\documentclass[11pt,a4paper]{article}
\usepackage{jheppub}

\def\beq{\begin{equation}}
\def\eeq{\end{equation}}

\usepackage{amsmath}
\usepackage{epsfig,epsf}
\usepackage{graphicx}
\usepackage{verbatim}

\title{NLO  JIMWLK  evolution unabridged}
\author[a]{Alex Kovner,}
\author[b]{Michael Lublinsky,} 
\author[b]{and Yair Mulian}

\affiliation[a]{Physics Department, University of Connecticut, 2152 Hillside road, Storrs, CT 06269, USA}
\affiliation[b]{Physics Department, Ben-Gurion University of the Negev, Beer Sheva 84105, Israel}

\usepackage{amsmath}
\usepackage{epsfig,epsf}
\usepackage{graphicx}
\usepackage{verbatim}
\usepackage{epstopdf}

\def\beq{\begin{equation}}
\def\eeq{\end{equation}}

\date{\today}
\abstract{
In Ref. \cite{nlojimwlk} we presented the  JIMWLK Hamiltonian for high energy evolution of 
QCD amplitudes  at the next-to-leading order accuracy in $\alpha_s$.
In the present paper we provide details of our original derivation, which was not reported in \cite{nlojimwlk}, 
and provide the Hamiltonian in the form appropriate for action on color singlet as well as color nonsinglet states.
The   rapidity evolution of the quark dipole   generated by this Hamiltonian is computed and compared 
with the corresponding  result of Balitsky and Chirilli \cite{BC}.  
We then establish the equivalence between the NLO JIMWLK Hamiltonian  and 
the NLO version of the Balitsky's hierarchy \cite{BClast}, which includes action on nonsinglet combinations of Wilson lines.
Finally, we present complete evolution equation for  three-quark Wilson loop operator, 
thus extending the results of Grabovsky \cite{Grab}.  
 }

\begin{document}
\maketitle

\newpage



\section{Introduction and Conclusions}

The JIMWLK Hamiltonian \cite{jimwlk} is the limit of the QCD Reggeon Field Theory (RFT),
applicable for computations of high energy  scattering amplitudes   of dilute (small parton number) projectiles on dense (nuclei) targets.  In general it predicts rapidity evolution of any hadronic observable $\cal O$ via the functional equation of the form
\beq\label{1}
\frac{d}{dY}\,{\cal O} \,=\,-\,H^{JIMWLK}\,{\cal O}\,.
\eeq
In ref. \cite{jimwlk}, the JIMWLK Hamiltonian was derived in the leading order in $\alpha_s$ in pQCD. It  contains a wealth of information about high energy evolution equations. In the dilute-dilute limit it reproduces  the linear  BFKL equation \cite{bfkl} and
its BKP extension \cite{bkp}. Beyond the dilute limit, the Hamiltonian incorporates non-linear effects responsible for unitarization of scattering amplitudes.
For practical applications,   the JIMWLK  evolution is usually replaced by the Balitsky-Kovchegov  (BK) non-linear evolution equation \cite{Bal,KOV}, which at large $N_c$ describes the growth of the  gluon density with energy and the gluon saturation phenomenon \cite{GLR} as reflected in the evolution of the scattering amplitude of a fundamental dipole $s$
\begin{equation}s(x,y)=\frac{1}{N_c}tr[S(x)\,S^\dagger(y)]\ .
\end{equation} 
The Wilson line
$S(x)$, in the high energy eikonal approximation represents the scattering amplitude of a quark at the transverse coordinate $x$.  
There exist numerous phenomenological applications  of the BK equation to DIS, heavy ion 
collisions and proton-proton collisions at the LHC \cite{review}.  Successful BK phenomenology mandates inclusion of next to leading order corrections, since at leading order the evolution predicted by the BK equation is too rapid to describe experimental data. Currently only the running coupling corrections are included in applications, although it is clearly desirable to include all next to leading corrections. 

The complete set of such corrections to the evolution of a fundamental dipole was calculated by Balitsky and Chirilli \cite{BC}, following on the earlier works \cite{BalNLO}. 
This result generalizes the NLO  BFKL equation  \cite{NLOBFKL} and reduces to it in the linearized approximation. Grabovsky \cite{Grab} computed certain, connected,
parts of the NLO evolution equation for three-quark  Wilson loop operator in the $SU(3)$ theory (which we will sometimes refer to as "baryon")
\begin{equation} B\equiv\epsilon_{ijk}\epsilon_{lmn}\,S^{il}(u)S^{jm}(v)S^{kn}(w)\ .
\end{equation}
Projected on  the charge conjugation odd sector, the  operator $B$
is related to the odderon, which  at NLO was independently studied  in \cite{NLOBKP}.

The NLO extension of the JIMWLK framework is imperative for calculation of more general amplitudes, beyond the dipole, 
which determine interesting experimental observables like single- and double inclusive particle production.  
Thanks to the above mentioned  major progress in the NLO computations, 
in \cite{nlojimwlk} we have presented the NLO JIMWLK Hamiltonian which reproduces these results by simple algebraic application to the relevant amplitudes.
Ref.\cite{nlojimwlk}  appeared simultaneously with \cite{BClast}, which directly calculated  many elements of the general 
Balitsky's hierarchy at NLO.  Our construction in \cite{nlojimwlk} was based upon  two major pieces of input. First, the general form
of the NLO JIMWLK Hamiltonian was deduced from the hadronic  wave-function computation in the light cone perturbation
theory \cite{LM}. This allowed us to parametrize the Hamiltonian in terms of only five kernels. These kernels were then fully reconstructed by comparing
the evolution generated by the Hamiltonian with the detailed results of \cite{BC} and \cite{Grab}. 

Using a similar strategy, in ref. \cite{KLMconf} (see also ref. \cite{Simon}) we have constructed the NLO JIMWLK Hamiltonian  for ${\cal N}=4$ SUSY, which is a conformal field theory. The question
addressed in \cite{KLMconf} was   whether  the conformal invariance of the theory is preserved on the level of the effective RFT. While
the leading order JIMWLK equation is conformally invariant when applied on gauge invariant states,  
the NLO evolution of the color dipole derived in \cite{BC} as well as the explicit form of the NLO JIMWLK Hamiltonian given in \cite{nlojimwlk} 
naively appear to violate conformal invariance.  The origin of this seeming violation lies in the fact that 
 these  NLO calculations involve hard cutoff in rapidity space, which itself is not conformally invariant \cite{FF}. 
It was shown in \cite{N=4} that in the particular case of the dipole evolution it is possible to redefine the dipole operator in such a way that its evolution becomes conformally invariant.  In \cite{KLMconf} we showed that the NLO JIMWLK equation for ${\cal N}=4$ theory in fact does have  exact conformal invariance, even though it is derived with sharp rapidity cutoff. The conformal transformation of the Wilson line operators is  different from the naive one.  We were able to 
construct perturbatively the conformal symmetry representation on the space of Wilson lines.  The modified transformation was found to be an 
exact (up to NNLO terms) symmetry of the Hamiltonian. 





 

The present paper continues our study of the NLO JIMWLK Hamiltonian, as well as provides detailed derivation of the results presented in \cite{nlojimwlk}. In Section 2 we provide a quick overview of the JIMWLK formalism and, following  
\cite{nlojimwlk}, present  the NLO Hamiltonian and sketch the path to its derivation.  While in \cite{nlojimwlk} we have presented the final form of the Hamiltonian, we did not provide the details of the derivation. One of the purposes of the present paper is to rectify this situation. We apply the Hamiltonian, parametrized by the five kernels  to the dipole operator $s$.
The resulting evolution equation
for  $s$ is then confronted with the explicit NLO calculation of \cite{BC}, which is quoted in eq.(\ref{dipole}) for self-consistency of presentation.  Matching 
various terms in the evolution we are able to fix (most of) the kernels.  This is a straightforward but
rather lengthy computation most of which is contained in  Appendix A.

In Section 3, we  generalize the Hamiltonian so that it generates correct evolution also for color nonsinglet operators. 
The NLO JIMWLK Hamiltonian of \cite{nlojimwlk} was constructed so that it generates unambiguous
 evolution of gauge invariant operators only, that is operators invariant under the action of the $SU_L(N_c)\times SU_R(N_c)$ group.  While most of the operators of physical interest are of this type, some color nonsinglet operators  have also been a focus of attention in the context of high energy evolution. The prime example is the operator representing the Reggeized gluon, which played a very important role in the development of high energy evolution ideas, especially in the perturbative domain. Another example of an interesting nonsinglet observable, is inclusive gluon production amplitude, which is invariant only under the vector subgroup of $SU_L(N_c)\times SU_R(N_c)$ \cite{tolga}.
 Recently ref. \cite{BClast} presented evolution of one, two, and three Wilson lines with uncontracted  color indices. Using these results we deduce the NLO JIMWLK Hamiltonian valid on the entire Hilbert space of RFT, which includes nonsinglet operators.
The action of the generalized Hamiltonian  on any singlet operator, 
 is equivalent to that of the original Hamiltonian presented in \cite{nlojimwlk}. On the other hand, when applied to one, two, and three
Wilson lines with uncontracted color indices, it reproduces the results of \cite{BClast}. As a by product of this calculation we prove mutual consistency between \cite{nlojimwlk} and \cite{BClast}, which appeared simultaneously and have not been directly compared with each other so far.

In Section 4, we apply the Hamiltonian to the three-quark singlet operator $B$ and derive its complete evolution equation.  
Comparison of the fully connected part of this calculation with the results of \cite{Grab}, was part of the input which allowed us  in \cite{nlojimwlk} to deteremine two kernels in the Hamiltonian. The remaining terms in the evolution of $B$ presented in the current paper are new.  
They demonstrate the power of the Hamiltonian method, as no additional NLO calculations are needed to be performed to derive the evolution of $B$ or any other observable, apart from those that are necessary to determine the Hamiltonian.
In our previous paper \cite{KLMconf}, a conformal extension ${\cal B}$ of the operator $B$ was constructed. The concluding part  of Section 4  
discusses the evolution equation for the operator $\cal B$ in ${\cal N}=4$ theory. By construction, the resulting evolution is conformally invariant.
This part of our work overlaps with ref. \cite{BalGrab}, which is being released concurrently with the present paper.

\section{JIMWLK Hamiltonian at Leading and Next to Leading orders.}
The JIMWLK Hamiltonian defines a two-dimensional non-local field theory of a  unitary matrix (Wilson line) $S(x)$. 
The leading order  Hamiltonian is:
\begin{eqnarray}\label{LO}
H^{LO\ JIMWLK}= \int _{z,x,y} K^{LO}_{x,y,z} \left[ J^a_L(x)J^a_L(y)+J_R^a(x)J_R^a(y)- 2J_L^a(x)S^{ab}_A(z)J^b_R(y)\right]\,.
\end{eqnarray}
The left and right $SU(N_c)$ rotation generators, when acting on functions of $S$ have the representation
\begin{eqnarray}\label{LR}
J^a_L(x)=tr\left[\frac{\delta}{\delta S^{T}_x}t^aS_x\right]-tr\left[\frac{\delta}{\delta S^{*}_x}S^\dagger_xt^a\right] ; \ 
J^a_R(x)=tr\left[\frac{\delta}{\delta S^{T}_x}S_xt^a\right] -tr\left[\frac{\delta}{\delta S^{*}_x}t^aS^\dagger_x\right]\,.
\end{eqnarray}
Here $t^a$ are $SU(N_c)$ generators in the fundamental representation, while $S_A$ is a unitary matrix in the adjoint representation - the gluon scattering amplitude. 
\begin{equation}\begin{split}
&J_{L}^{a}(x)S^{ij}(y)=(t^aS(x))^{ij}\delta(x-y)\,; \ \ \ \ \ \ \ \  \ \ \ \ \ \ \ J_{R}^{a}(x)S^{ij}(y)=(S(x)t^a)^{ij}\delta(x-y)\\
&J_{L}^{a}(x)=S_{A}^{ab}(x)J_{R}^{b}(x)\,; \ \ \ \ \ \ \ \  \ \ \ \ \ \ \  \ \ \ \  \ \ \ \ \ \ \ \ \ J_{R}^{a}(x)=S_{A}^{ba}(x)J_{L}^{b}(x)\,. \\
 \end{split}\end{equation}
The leading order kernel  is given by
\beq
K^{LO}(x,y,z)\,=\,\,\frac{\alpha_s}{2\,\pi^2}\,\frac{X\cdot Y}{ X^2\,Y^2}\,.
\eeq
We use the notations of ref. \cite{BC} $X\equiv x-z$,  $X^\prime\equiv x-z^\prime$, $Y\equiv y-z$,    $ Y^\prime \equiv y-z^\prime$,
$W\equiv w-z$,  and  $ W^\prime \equiv w-z^\prime$.

The LO Hamiltonian is invariant under $SU_L(N_c)\times SU_R(N_c)$ rotations, which reflects gauge invariance of scattering amplitudes.
When acting on gauge invariant operators (operators invariant separately under $SU_L(N_c)$ and $SU_R(N_c)$ rotations), the kernel $K^{LO}$ can be 
substituted by the so called dipole kernel
\begin{equation}
K^{LO}(x,y,z)\ \rightarrow \ - \frac{1}{2}M(x,y;z); \ \ \ \ \ \ \ \ M(x,y;z)\,=\,\frac{\alpha_s}{2\,\pi^2}\,\frac{(x-y)^2}{X^2\,Y^2}
\end{equation}
which vanishes at $x=y$ and has a better IR behavior. In addition, the Hamiltonian is invariant under the $Z_2$ transformation $S\rightarrow S^\dagger; \ \ J_L\rightarrow -J_R$, which in \cite{reggeon} was identified as signature, and the charge conjugation symmetry $S\rightarrow S^*$.

The JIMWLK Hamiltonian is derivable from perturbatively computable hadronic wavefunction \cite{KL}. At LO, the wavefunction schematically (omitting transverse coordinates and color indices)  has the form
\beq
|\psi\rangle\,=\, (1\,-\,g_s^2\,\kappa_0\, JJ)\,|\,no\,soft\, gluons \rangle \,+\,g_s \kappa_1\,J\,|\,one\, soft \,gluon\rangle\,.
\eeq
Here $J$ is the color charge density (of valence gluons) which emits the  soft gluons at the next step of the evolution. The probability amplitude for a single gluon emission $\kappa_1$ is essentially the
Weizsacker-Williams field. A sharp cutoff in longitudinal momenta is implied in the separation between valence and soft modes in the wavefunction.
The $\kappa_0\, JJ$ term is due to normalization of the wavefunction at the order $g^2_s$, $\kappa_0\sim \kappa_1^2$.  
The JIMWLK Hamiltonian is obtained by computing the expectation value of the $\hat S$-matrix operator (expanded to first order in longitudinal phase space):
\beq
H^{JIMWLK}\,=\,\langle \psi|\, \hat S \,-\,1\,|\psi\rangle
\eeq
The fact that  the real term ($JSJ$) and the virtual term ($JJ$) emerge with the very same kernel $K^{LO}$  in eq. (\ref{LO}) is a direct consequence of the
wavefunction normalization. Note that $H^{JIMWLK}$ vanishes if we set $S(z)=1$ and $J_L=J_R$. This property reflects the fact that if none of the particles in the wave function scatter, the scattering matrix does not evolve with energy. This fundamental property must be preserved also at NLO.

To compute the NLO Hamiltonian,  the wavefunction has to be computed to order $g_s^3$ and normalized to order $g_s^4$ \cite{LM}. Each emission off the valence gluons in the wave function brings a factor of color charge density $J$. At NLO at most two soft gluons can be emitted, and therefore the general form of the wave function at NLO is:
 \begin{eqnarray}\label{psi}
|\psi\rangle&=& (1\,-\,g_s^2\,\kappa_0\, JJ\,-\,g_s^4(\delta_1\,JJ\,+\,\delta_2\,JJJ\,+\,\delta_3\,JJJJ)\,|\,no\,soft\, gluons \rangle +
\nonumber \\
&+&(\,g_s \kappa_1\,J\,+\,g_s^3\epsilon_1\, J\,+\, g_s^3\,\epsilon_2\,J \,J)\,|\,one\, soft \,gluon\rangle
+g_s^2 (\epsilon_3\, J\,\,+\,\epsilon_4\, JJ)\,|\,two\, soft \,gluons\rangle\nonumber \\
&+&g_s^2\,\epsilon_5\,J\,|\,quark-antiquark\rangle\,.
\end{eqnarray}
More constraints on the form of the Hamiltonian come from the symmetries of the theory.
As discussed in detail in \cite{reggeon}, the theory must have $SU_L(N)\times SU_R(N)$ symmetry, which in QCD terms is the gauge symmetry of $|in\rangle$ and 
$|out\rangle$ states and two discrete symmetries: the charge conjugation $S(x)\rightarrow S^*(x)$, and another $Z_2$ symmetry:  $S(x)\rightarrow S^\dagger(x)$,  $J^a_L(x)\leftrightarrow -J^a_R(x)$ which in \cite{reggeon} was identified with signature, and can be understood as the combination of charge conjugation and time reversal symmetry \cite{iancutri}.

The algorithm of obtaining the Hamiltonian for high energy evolution starting from the soft gluon wave function has been described in detail in \cite{KL}. 
Given the general form eq.(\ref{psi}) and the symmetry constraints, the Hamiltonian can be parametrized in terms of six kernels
\begin{eqnarray}\label{NLO1}
&&H^{NLO\ JIMWLK}= \int_{x,y,z} K_{JSJ}(x,y;z) \left[ J^a_L(x)J^a_L(y)+J_R^a(x)J_R^a(y)-2J_L^a(x)S_A^{ab}(z)J^b_R(y)\right]  \nonumber \\
 &&+\int_{x\,y\, z\,z^\prime}K_{JSSJ}(x,y;z,z^\prime)\left[f^{abc}f^{def}J_L^a(x) S^{be}_A(z)S^{cf}_A(z^\prime)J_R^d(y)- N_c J_L^a(x)S^{ab}_A(z)J^b_R(y)\right] 
 \nonumber \\
 &&+\int_{x,y, z,z^\prime} K_{q\bar q}(x,y;z,z^\prime)\left[2\,J_L^a(x) \,tr[S^\dagger(z)\, t^a\,S (z^\prime)t^b]\,J_R^b(y)\,
 -\, J_L^a(x)\,S^{ab}_A(z)\,J^b_R(y)\right] \nonumber \\
 &&+\int_{w,x,y, z,z^\prime}K_{JJSSJ}(w;x,y;z,z^\prime)f^{acb}\,\Big[J_L^d(x)\, J_L^e(y)\, S^{dc}_A(z)\,S^{eb}_A(z^\prime)\,J_R^a(w)\nonumber \\
 &&-\,J_L^a(w)\,S^{cd}_A(z)\,S^{be}_A (z^\prime)\,J_R^d(x)\,J_R^e(y)\,\Big] \nonumber \\
 &&+\int_{w,x,y, z}\, K_{JJSJ}(w;x,y;z)\,f^{bde}\,\Big[  J_L^d(x) \,J_L^e(y) \,S^{ba}_A(z)\,J_R^a(w)\,-\, 
J_L^a(w)\,S^{ab}_A(z)\,J_R^d(x)\,J_R^e(y)\Big]  \,  \nonumber \\
 &&+\int_{w,x,y}\, K_{JJJ}(w;x,y)\,[J_L^d(x) \,J_L^e(y) \,J_L^b(w)\,-\, 
J_R^d(x)\,J_R^e(y)\,J_R^b(w)]\,.
\end{eqnarray}
The ordering of various factors in eq.(\ref{NLO1}) is such that all factors of $J$ are assumed to be to the right of all factors of $S$, and therefore $J$'s do not act on $S$ in the Hamiltonian. The ordering of the different factors of $J$ between themselves is important, since the operators $J$ do not commute with each other. Throughout the calculation this ordering is kept as explicitly indicated in eq.(\ref{NLO1}).

No other color structures appear in the light cone wave function calculation. The discrete symmetries require the kernels $K_{JSSJ}$ and $K_{q\bar q}$
to be symmetric under the interchanges $z\leftrightarrow z^\prime$ or $x\leftrightarrow y$, while  $K_{JJSSJ}$ to be antisymmetric under simultaneous 
 interchange $z\leftrightarrow z^\prime$ and  $x\leftrightarrow y$. 
 
 Our strategy is to fix the various kernels by calculating the action of the Hamiltonian on the dipole, and the baryon operator. 
 As it turns out, this calculation is sufficient to determine all but one kernels. The last virtual term vanishes when acting on both, the dipole and the baryon operator, and thus one needs additional information to determine the kernel $K_{JJJ}$. One way of dealing with it is to consider the action of the Hamiltonian on nonsinglet combinations of Wilson lines, where in general it gives a nonvanishing contribution. Another way is to use conformal invariance of tree level QCD \cite{KLMconf}. At the next to leading order the only violation of conformal invariance in QCD should come from the running coupling constant. The terms associate with the running constant are the second and third terms in eq.(\ref{NLO1}). 
 As we have shown in \cite{KLMconf}, the requirement of conformal invariance of the rest of the Hamiltonian determines $K_{JJJ}$ in terms of other kernels. Using the results of  
 \cite{KLMconf}, we can therefore further restrict the ansatz for the Hamiltonian, and write it in terms of five kernels: 
\begin{eqnarray}\label{NLO}
&&H^{NLO\ JIMWLK}= \int_{x,y,z} K_{JSJ}(x,y;z) \left[ J^a_L(x)J^a_L(y)+J_R^a(x)J_R^a(y)-2J_L^a(x)S_A^{ab}(z)J^b_R(y)\right]  \nonumber \\
 &&+\int_{x\,y\, z\,z^\prime}K_{JSSJ}(x,y;z,z^\prime)\left[f^{abc}f^{def}J_L^a(x) S^{be}_A(z)S^{cf}_A(z^\prime)J_R^d(y)- N_c J_L^a(x)S^{ab}_A(z)J^b_R(y)\right] 
 \nonumber \\
 &&+\int_{x,y, z,z^\prime} K_{q\bar q}(x,y;z,z^\prime)\left[2\,J_L^a(x) \,tr[S^\dagger(z)\, t^a\,S (z^\prime)t^b]\,J_R^b(y)\,
 -\, J_L^a(x)\,S^{ab}_A(z)\,J^b_R(y)\right] \nonumber \\
 &&+\int_{w,x,y, z,z^\prime}K_{JJSSJ}(w;x,y;z,z^\prime)f^{acb}\,\Big[J_L^d(x)\, J_L^e(y)\, S^{dc}_A(z)\,S^{eb}_A(z^\prime)\,J_R^a(w)\nonumber \\
 &&-\,J_L^a(w)\,S^{cd}_A(z)\,S^{be}_A (z^\prime)\,J_R^d(x)\,J_R^e(y)\, +\,\frac{1}{3}[J_L^c(x) \,J_L^b(y) \,J_L^a(w)\,-\, 
J_R^c(x) \,J_R^b(y)\,J_R^a(w)]\,\Big] \nonumber \\
 &&+\int_{w,x,y, z}\, K_{JJSJ}(w;x,y;z)\,f^{bde}\,\Big[  J_L^d(x) \,J_L^e(y) \,S^{ba}_A(z)\,J_R^a(w)\,-\, 
J_L^a(w)\,S^{ab}_A(z)\,J_R^d(x)\,J_R^e(y) \,  \nonumber \\
 &&+\,\frac{1}{3}[J_L^d(x) \,J_L^e(y) \,J_L^b(w)\,-\, 
J_R^d(x)\,J_R^e(y)\,J_R^b(w)]
\Big] \,.
\end{eqnarray}

 The kernels $K_{JJSJ}$ and $K_{JJSSJ}$  are now fixed by acting with the Hamiltonian on the operator $B$ and comparing the result to that of ref. \cite{Grab}.
 This calculation will be presented in Section 4. The other  kernels are then determined by acting on the dipole $s$  and
comparing with results of \cite{BC}. The details of this calculation  are presented in the  Appendix A.

The resulting expressions for the kernels are: 
\begin{eqnarray}
&&K_{JJSSJ}(w;x,y;z,z^\prime)=-i
\frac{\alpha_s^2}{ 2\,\pi^4}
\left(\frac{X_iY^\prime_j}{ X^2Y^{\prime 2}}
\right)
\nonumber \\
&&\ \ \ \ \ \  \ \ \ \ \ \ \ \times \Big(\frac{\delta_{ij}}{2 (z-z^\prime)^2}+\frac{(z^\prime-z)_i W^\prime_j}{ (z^\prime-z)^2 W^{\prime 2}}+
\frac{(z-z^\prime)_j W_i}{ (z-z^\prime)^2 W^{ 2}}-\frac{W_i W^\prime_j}{ W^2 W^{\prime 2}}
\Big)\ln\frac{W^2}{ {W'}^2}
\end{eqnarray}
\begin{eqnarray}\label{KJJSJ}
K_{JJSJ}(w;x,y;z)\,=\,-\,i\,\frac{\alpha_s^2}{ 4\, \pi^3 }\,\Big[ \frac{X\cdot W}{ X^2\,W^2}\,-\, \frac{Y\cdot W}{ Y^2\,W^2}    \Big] \ln\frac{Y^2}{ (x-y)^2}\,\ln\frac{X^2}{ (x-y)^2},
\end{eqnarray}
\begin{eqnarray}\label{Kqq}
K_{q\bar q}(x,y;z,z^\prime) =-\frac{\alpha_s^2\,n_f}{ 8\,\pi^4}
\Big\{
\frac{{X'}^2Y^2+{Y'}^2X^2-(x-y)^2(z-z')^2}{ (z-z')^4(X^2{Y'}^2-{X'}^2Y^2)}
\ln\frac{X^2{Y'}^2}{ {X'}^2Y^2}-\frac{2}{(z-z^\prime)^4}\Big\},\nonumber \\
\end{eqnarray}
Defining for convenience
\begin{eqnarray}
\tilde K(x,y,z,z^\prime)\,=\frac{i}{2}\,\Big[K_{JJSSJ}(x;x,y;z,z^\prime)&-&K_{JJSSJ}(y;x,y;z,z^\prime)-K_{JJSSJ}(x;y,x;z,z^\prime)\nonumber \\
&+&K_{JJSSJ}(y;y,x;z,z^\prime)\Big]\ ,
\end{eqnarray}
we can write the remaining kernels as:
\begin{eqnarray}\label{KJSSJ}
&&K_{JSSJ}(x,y;z,z^\prime) = \frac{\alpha_s^2}{16\,\pi^4}
\Bigg[\,-\,\frac{4}{ (z-z^\prime)^4}\,+\,
\Big\{2\frac{X^2{Y'}^2+{X'}^2Y^2-4(x-y)^2(z-z')^2}{ (z-z')^4[X^2{Y'}^2-{X'}^2Y^2]}\nonumber\\ 
&&+
~\frac{(x-y)^4}{ X^2{Y'}^2-{X'}^2Y^2}\Big[
\frac{1}{X^2{Y'}^2}+\frac{1}{ Y^2{X'}^2}\Big]
+\frac{(x-y)^2}{(z-z')^2}\Big[\frac{1}{X^2{Y'}^2}-\frac{1}{ {X'}^2Y^2}\Big]\Big\}
\ln\frac{X^2{Y'}^2}{ {X'}^2Y^2}\Bigg]\nonumber \\ 
&&+\,\tilde K(x,y,z,z^\prime)\label{tildeK}.
\end{eqnarray}
\begin{eqnarray}\label{KJSJ}
K_{JSJ}(x,y;z) =&-&\frac{\alpha_s^2}{16 \pi^3}
\frac{(x-y)^2}{X^2 Y^2}\Big[b\ln(x-y)^2\mu^2
-b\frac{X^2-Y^2}{ (x-y)^2}\ln\frac{X^2}{Y^2}+
(\frac{67}{9}-\frac{\pi^2}{ 3})N_c-\frac{10}{ 9}n_f\Big]\nonumber \\
&-& \frac{N_c}{2}\ \int_{z^\prime}\, \tilde K(x,y,z,z^\prime).
\end{eqnarray}
Here $\mu$ is the normalization point in the $\overline{MS}$ scheme and
$b=\frac{11}{3}N_c-\frac{2}{3}n_f$ is the first coefficient of the $\beta$-function.
The kernels satisfy the following useful identities \cite{BClast}:
\begin{eqnarray}\label{Krelate}
&&K_{JJSJ}(w;x,y;z)=
  \int_{z^\prime}\, \left[K_{JJSSJ}(y;w,x;z,z^\prime)-K_{JJSSJ}(x;w,y;z,z^\prime)\right],\nonumber \\
&&\int_{z^\prime}K_{JJSSJ}(y,x,y,z,z^{\prime})=0,\nonumber \\
 && \int_{z}K_{JJSJ}(y,x,y,z)=\int_{z,z^{\prime}}K_{JJSSJ}(y,y,x,z,z^{\prime})=0,\nonumber \\
&& \int_{z^{\prime},z}\widetilde{K}(x,y,z,z^{\prime})=i\, \int_z\left[K_{JJSJ}(y,x,y,z)+K_{JJSJ}(x,y,x,z)\right]=0\,.
\end{eqnarray}
Since the above determination of the kernels relies only on the action of the Hamiltonian on color singlet operators, the kernels 
are determined only modulo terms that do not depend on (at least) one of the coordinates carried by one of the charge density operators $J$. 
One can add to the Hamiltonian an arbitrary operator proportional to $Q_{L(R)}^a=\int d^2x J_{L(R)}^a(x)$ without altering its action on singlets, since $Q^a_{L(R)}$ annihilates any color singlet state.  The terms proportional to $1/ (z-z^\prime)^4$ and independent of $X$ and $Y$ in $K_{JSSJ}$ and $K_{q\bar q}$  are  examples of such terms. We assigned them to the kernels in this form, so that the kernels vanish at $x=y$ analogously to the dipole kernel at LO.  

As long as one is interested in color singlet operators only, the above form of the Hamiltonian is perfectly adequate. However some interesting observables, like single gluon inclusive production require the knowledge of observables which are singlets only under the vector subgroup of $SU_L(N_c)\times SU_R(N_c)$\cite{tolga}. It is thus useful to generalize the Hamiltonian so that it generates correct evolution of such operators.
This requires additional input. While the structure of the Hamiltonian given in eq.(\ref{NLO1}) remains valid in the general case, some of the kernels have to be modified by additional terms which do not depend on one of the coordinates of a charge density operator $J^a$.
These additional terms can be inferred by considering the action of the Hamiltonian on nonsinglet products of Wilson line and comparing the results to \cite{BClast}.  This is the subject of the next section.


\section{NLO JIMWLK for color non-singlet operators. Comparison with \cite{BClast} }


In this section we generalize the NLO JIMWLK Hamiltonian to make it applicable to color non-singlets. The generalization is done basically "by inspection". 
We modify the kernels in the ansatz
(\ref{NLO}) and show that the resulting Hamiltonian reproduces the results of \cite{BClast}. The modified kernels are:
\begin{equation}\begin{split}
&K_{JSJ}(x,y,z)\rightarrow
\bar{K}_{JSJ}(x,y,z) \equiv K_{JSJ}(x,y,z)+\\
&+{\alpha_s^2\over 16\pi^3}\left\{\left[\frac{1}{X^{2}}+\frac{1}{Y^{2}}\right]\left[\left(\frac{67}{9}-\frac{\pi^{2}}{3}\right)N_{c}-\frac{10}{9}n_{f}\right] \right. 
\left. +\frac{b}{X^{2}}\ln X^{2}\mu^{2}+\frac{b}{Y^{2}}\ln Y^{2}\mu^{2}\right\};
 \end{split}\end{equation}
 \begin{equation}\begin{split}
K_{JSSJ}(x,y;z,z^{\prime})\rightarrow\bar{K}_{JSSJ}(x,y;z,z^{\prime})&\equiv K_{JSSJ}(x,y;z,z^{\prime})\\
&+\frac{\alpha_{s}^{2}}{8\pi^{4}}\left[\frac{4}{(z-z^{\prime})^{4}}-\frac{I(x,z,z^{\prime})}{(z-z^{\prime})^{2}}-\frac{I(y,z,z^{\prime})}{(z-z^{\prime})^{2}}\right];\\ 
\end{split}\end{equation}
\begin{equation}\begin{split}
&K_{q\bar{q}}(x,y;z,z^{\prime})\rightarrow \bar{K}_{q\bar{q}}(x,y;z,z^{\prime})\equiv K_{q\bar{q}}(x,y;z,z^{\prime})-\frac{\alpha_{s}^{2}n_{f}}{8\pi^{4}}\left[\frac{I_{f}(x,z,z^{\prime})}{(z-z^{\prime})^{2}}+\frac{I_{f}(y,z,z^{\prime})}{(z-z^{\prime})^{2}}\right], \\
\end{split}\end{equation}
where $I$ and $I_f$ are defined as in \cite{BClast}:
\begin{equation}\begin{split}
&I(x,z,z^{\prime})=\frac{1}{X^{2}-(X^{\prime})^{2}}\ln\frac{X^{2}}{(X^{\prime})^{2}}\left[\frac{X^{2}+(X^{\prime})^{2}}{(z-z^{\prime})^{2}}-\frac{X\cdot X^{\prime}}{X^{2}}-\frac{X\cdot X^{\prime}}{(X^{\prime})^{2}}-2\right];\\
&I_{f}(x,z,z^{\prime})=\frac{2}{(z-z^{\prime})^{2}}-\frac{2X\cdot X^{\prime}}{(z-z^{\prime})^{2}(X^{2}-(X^{\prime})^{2})}\ln\frac{X^{2}}{(X^{\prime})^{2}}.\\
\end{split}\end{equation}
The kernels $K_{JJSJ}$  and $K_{JJSSJ}$ remain unchanged.

As discussed above, the modifications of the kernels involve only extra terms  that are independent of either $x$ or $y$. These additional terms  do not contribute to evolution of 
gauge invariant operators.  

 Following \cite{BClast} we start by considering the evolution of a single fundamental Wilson line $(S(x))_{ij}$. We then proceed to act on a product of two Wilson lines with uncontracted indices, $(S(x))_{ij}(S(y))_{kl}$,
 and finally on a product of three, $(S(x))_{ij}(S(y))_{kl}(S(w))_{mn}$. In order to compare with \cite{BClast}, we only need to consider connected terms, that is the terms in which each factor of $S$ in the operator is acted upon at least one charge density operator $J$ in the Hamiltonian.


\subsection{Self-interaction}
The evolution equation of a single Wilson line $S(x)_{ij}$ as calculated in \cite{BClast}, is:
\begin{equation}\begin{split}\label{S1}
&\frac{d}{dY}(S(x))_{ij}=\\
&=\frac{\alpha_{s}^{2}}{4\pi^{3}}\int_{z}\frac{1}{X^{2}}\left[\left[\frac{11N_{c}}{3}-\frac{2}{3}n_{f}\right]\ln X^{2}\mu^{2}+\frac{67N_{c}}{9}-\frac{\pi^{2}N_{c}}{3}-\frac{10}{9}n_{f}\right]\\
&\times \left[S_{A}^{ab}(z)-S_{A}^{ab}(x)\right](t^{a}S(x)t^{b})_{ij}\\
&+\frac{\alpha_{s}^{2}}{8\pi^{4}}\int_{z,z^{\prime}}S_{A}^{dd^{\prime}}(z)\left[S_{A}^{ee^{\prime}}(z^{\prime})-S_{A}^{ee^{\prime}}(z)\right]f^{ade}f^{bd^{\prime}e^{\prime}}
\left[\frac{2I(x,z,z^{\prime})}{(z-z^{\prime})^{2}}-\frac{4}{(z-z^{\prime})^{4}}\right](t^{a}S(x)t^{b})_{ij}\\
&+\frac{n_{f}\alpha_{s}^{2}}{4\pi^{4}}\int_{z,z^{\prime}}\frac{2I_{f}(x,z,z^{\prime})}{(z-z^{\prime})^{2}}(t^{a}S(x)t^{b})_{ij}tr\left[t^{a}S(z)t^{b}(S^{\dagger}(z^{\prime})-S^{\dagger}(z))\right]\\
&+\frac{i\alpha_{s}^{2}}{8\pi^{4}}\int_{z,z^{\prime}}S_{A}^{dd^{\prime}}(z)\left[S_{A}^{ee^{\prime}}(z^{\prime})-S_{A}^{ee^{\prime}}(z)\right]\frac{X\cdot X^{\prime}}{(z-z^{\prime})^{2}X^{2}(X^{\prime})^{2}}\ln\frac{X^{2}}{(X^{\prime})^{2}}\\
&\times \left[f^{ad^{\prime}e^{\prime}}(\{t^{d},t^{e}\}S(x)t^{a})_{ij}-f^{ade}(t^{a}S(x)\{t^{d^{\prime}},t^{e^{\prime}}\})_{ij}\right]\,.\\
\end{split}\end{equation}

Acting on $S(x)_{ij}$ with our modified Hamiltonian we obtain:
\begin{equation}\begin{split}
&\frac{d}{dY}(S(x))_{ij}=-H^{NLO\ JIMWLK}(S(x))_{ij}=\\
&=\int_{z}2\bar{K}_{JSJ}(x,x;z)\left[S_{A}^{ab}(z)-S_{A}^{ab}(x)\right](t^{a}S(x)t^{b})_{ij} \\
&+\int_{z,z^{\prime}}\left[K_{JSSJ}(x,x;z,z^{\prime})+\frac{\alpha_{s}^{2}}{8\pi^{4}}\left[\frac{4}{(z-z^{\prime})^{4}}-{2I(x,z,z^{\prime})\over (z-z^{\prime})^{2}}\right]\right]f^{adc}f^{bef}\\
&\times \left[S_{A}^{de}(z)\, S_{A}^{cf}(z)-S_{A}^{de}(z)\, S_{A}^{cf}(z^{\prime})\right](t^{a}S(x)t^{b})_{ij}\\
&-\int_{z,z^{\prime}}\left[K_{q\bar{q}}(x;x;z,z^{\prime})-\frac{\alpha_{s}^{2}n_{f}}{8\pi^{4}}\frac{2I_{f}(x,z,z^{\prime})}{(z-z^{\prime})^{2}}\right]\left[2tr\left[S^{\dagger}(z)t^{a}S(z^{\prime})t^{b}\right]-S_{A}^{ab}(z)\right](t^{a}S(x)t^{b})_{ij}\\
&+\int_{z,z^{\prime}}K_{JJSSJ}(x;x,x;z,z^{\prime})\left[S_{A}^{dc}(z)\, S_{A}^{eb}(z^{\prime})+\frac{1}{3}S_{A}^{dc}(x)\, S_{A}^{eb}(x)\right]\\
&\times \left[f^{ade}(t^{a}S(x)t^{c}t^{b})_{ij}-f^{acb}(t^{e}t^{d}S(x)t^{a})_{ij}\right]\\
&+\int_{z}K_{JJSJ}(x;x,x;z)\left[S_{A}^{ad}(z)+\frac{1}{3}S_{A}^{ad}(x)\right]\left[f^{cae}(t^{c}t^{e}S(x)t^{d})_{ij}-f^{cde}(t^{a}S(x)t^{e}t^{c})_{ij}\right]
 \end{split}\end{equation}
Since $K_{JSSJ}(x,x;z,z^{\prime})=K_{q\bar{q}}(x,x;z;z^{\prime})=K_{JJSJ}(x;x,x;z)=0$ we obtain:
\begin{equation}\begin{split}\label{S1we}
&\frac{d}{dY}(S(x))_{ij}=\\
&=\frac{\alpha_{s}^{2}}{4\pi^{3}}\int_{z}\left[\frac{b}{X^{2}}\ln X^{2}\mu^{2}+\frac{1}{X^{2}}\left[\left(\frac{67}{9}-\frac{\pi^{2}}{3}\right)N_{c}-\frac{10}{9}n_{f}\right]\right]\left[S_{A}^{ab}(z)-S_{A}^{ab}(x)\right](t^{a}S(x)t^{b})_{ij}\\
&+\frac{\alpha_{s}^{2}}{8\pi^{4}}\int_{z,z^{\prime}}\left[2I(x,z,z^{\prime})-\frac{4}{(z-z^{\prime})^{4}}\right]f^{adc}f^{bef}\left[S_{A}^{de}(z)\, S_{A}^{cf}(z^{\prime})-S_{A}^{de}(z)\, S_{A}^{cf}(z)\right](t^{a}S(x)t^{b})_{ij}\\
&+\frac{n_{f}\alpha_{s}^{2}}{2\pi^{4}}\int_{z,z^{\prime}}\frac{I_{f}(x,z,z^{\prime})}{(z-z^{\prime})^{2}}tr\left[\left(S^{\dagger}(z)-S^{\dagger}(z^{\prime})\right)t^{a}S(z^{\prime})t^{b}\right](t^{a}S(x)t^{b})_{ij}\\
&+\frac{i\alpha_{s}^{2}}{4\pi^{4}}\int_{z,z^{\prime}}\frac{X\cdot X^{\prime}}{(z-z^{\prime})^{2}X^{2}(X^{\prime})^{2}}\ln\frac{X^{2}}{(X^{\prime})^{2}}S_{A}^{dc}(z)\, S_{A}^{eb}(z^{\prime})\left[f^{acb}(t^{e}t^{d}S(x)t^{a})_{ij}-f^{ade}(t^{a}S(x)t^{c}t^{b})_{ij}\right]\,.\\
 \end{split}\end{equation}
 Notice, that while the kernel $K_{JJSSJ}(x;x,x;z,z^{\prime})$ is non-zero, the corresponding virtual term vanishes due
  to the anti-symmetry under exchange of $z$ and $z^{\prime}$.
Eq. (\ref{S1}) can be further simplified by using   (\ref{id1})
\begin{equation}\begin{split}
&\frac{i\alpha_{s}^{2}}{8\pi^{4}}\int_{z,z^{\prime}}S_{A}^{dd^{\prime}}(z)\left[S_{A}^{ee^{\prime}}(z^{\prime})-S_{A}^{ee^{\prime}}(z)\right]\frac{X\cdot X^{\prime}}{(z-z^{\prime})^{2}X^{2}(X^{\prime})^{2}}\ln\frac{X^{2}}{(X^{\prime})^{2}}\\
&\times \left[f^{ad^{\prime}e^{\prime}}(\{t^{d},t^{e}\}S(x)t^{a})_{ij}-f^{ade}(t^{a}S(x)\{t^{d^{\prime}},t^{e^{\prime}}\})_{ij}\right]\\
&=\frac{i\alpha_{s}^{2}}{4\pi^{4}}\int_{z,z^{\prime}}S_{A}^{dc}(z)S_{A}^{eb}(z^{\prime})\frac{X\cdot X^{\prime}}{(z-z^{\prime})^{2}X^{2}(X^{\prime})^{2}}\ln\frac{X^{2}}{(X^{\prime})^{2}}\left[f^{acb}(t^{e}t^{d}S(x)t^{a})_{ij}-f^{ade}(t^{a}S(x)t^{c}t^{b})_{ij}\right]\,.
 \end{split}\end{equation}
We then find that  eq.(\ref{S1}) and eq.(\ref{S1we}) are identical.
 
 \subsection{Pairwise interaction }
The evolution equation for the product of two Wilson lines derived in \cite{BClast},  is
\beq
\frac{d}{dY}(S(x))_{ij}(S(y))_{kl}\mid_{conn.}=\int_{z,z^\prime} \left({\cal A}_1+{\cal A}_2+{\cal A}_3\right) +\int_z \left( {\cal B}_1+{\cal B}_2\right)\,,
\eeq
when expressed in terms of the kernels we have defined above, 
\begin{eqnarray}
{\cal A}_1&=& 
\left[(t^{a}S(x))_{ij}(S(y)t^{b})_{kl}+(S(x)t^{b})_{ij}(t^{a}S(y))_{kl}\right] 
\nonumber \\
&\times&\Big\{S_{A}^{dd^{\prime}}(z)\left[S_{A}^{ee^{\prime}}(z^{\prime})-S_{A}^{ee^{\prime}}(z)\right] f^{ade}f^{bd^{\prime}e^{\prime}}\left (- \bar K_{JSSJ}(x,y,z,z^{\prime})+
\widetilde{K}(x,y,z,z^{\prime}\right)\nonumber \\
&-&2\bar K_{q\bar{q}}(x,y,z,z^{\prime}) tr\left[t^{a}S(z)t^{b}(S(z^{\prime})^{\dagger}-S(z)^{\dagger})\right]\Big\}\,;
\end{eqnarray}
\begin{eqnarray}
{\cal A}_2&=& {- 2}\,(S_{A}^{dd^{\prime}}(z)-S_{A}^{dd^{\prime}}(x))(S_{A}^{ee^{\prime}}(z^{\prime})-S_{A}^{ee^{\prime}}(y))\nonumber \\
&\times& \Big\{\left[f^{ad^{\prime}e^{\prime}}(t^{d}S(x)t^{a})_{ij}(t^{e}S(y))_{kl}-f^{ade}(t^{a}S(x)t^{d^{\prime}})_{ij}(S(y)t^{e^{\prime}})_{kl}\right]K_{JJSSJ}(x,x,y,z,z^{\prime})\nonumber \\
&+&\left[f^{ad^{\prime}e^{\prime}}(t^{d}S(x))_{ij}(t^{e}S(y)t^{a})_{kl}-f^{ade}(S(x)t^{d^{\prime}})_{ij}(t^{a}S(y)t^{e^{\prime}})_{kl}\right]K_{JJSSJ}(y,x,y,z,z^{\prime})\Big\}\,; \nonumber \\
\end{eqnarray}
\begin{eqnarray}
{\cal A}_3&=&-
S_{A}^{dd^{\prime}}(z)\Big\{ \left[f^{ad^{\prime}e^{\prime}}(S(x)t^{a})_{ij}(t^{d}t^{e}S(y))_{kl}-f^{ade}(t^{a}S(x))_{ij}(S(y)t^{e^{\prime}}t^{d^{\prime}})_{kl}\right]
\left(S_{A}^{ee^{\prime}}(z^{\prime})\right. \nonumber \\ 
&-&\left. S_{A}^{ee^{\prime}}(y)\right)
\left( K_{JJSSJ}(x,y,y,z,z^{\prime})+K_{JJSSJ}(y,y,x,z,z^{\prime})-K_{JJSSJ}(y,x,y,z,z^{\prime})\right)\nonumber \\
&+& \left[f^{ad^{\prime}e^{\prime}}(t^{d}t^{e}S(x))_{ij}(S(y)t^{a})_{kl}-f^{ade}(S(x)t^{e^{\prime}}t^{d^{\prime}})_{ij}(t^{a}S(y))_{kl}\right]\left(S_{A}^{ee^{\prime}}(z^{\prime})-S_{A}^{ee^{\prime}}(x)\right)\nonumber \\
&\times&\left( K_{JJSSJ}(y,x,x,z,z^{\prime})+K_{JJSSJ}(x,x,y,z,z^{\prime})-K_{JJSSJ}(x,y,x,z,z^{\prime})\right)\Big\}\,;
\end{eqnarray}
\begin{eqnarray}
{\cal B}_1&=& 
- \Big\{\left(S_{A}^{ab}(z)-S_{A}^{ab}(x)\right)\left[f^{bde}(t^{a}S(x)t^{d})_{ij}(S(y)t^{e})_{kl}-f^{ade}(t^{d}S(x)t^{b})_{ij}(t^{e}S(y))_{kl}\right]\nonumber \\
&\times& K_{JJSJ}(x,y,x,z)+\left(S_{A}^{ab}(z)-S_{A}^{ab}(y)\right)K_{JJSJ}(y,x,y,z)\nonumber \\
&\times& \left[f^{bde}(S(x)t^{e})_{ij}(t^{a}S(y)t^{d})_{kl}-f^{ade}(t^{e}S(x))_{ij}(t^{d}S(y)t^{b})_{kl}\right]
\Big\}\,;
\end{eqnarray}
\begin{eqnarray}
{\cal B}_2&=& 
\left[- \bar K_{JSJ}(x,y,z)-\frac{N_{c}}{2}\int_{z^{\prime}}\widetilde{K}(x,y,z,z^{\prime})\right]\left[S_{A}^{ab}(x)+S_{A}^{ab}(y)-2S_{A}^{ab}(z)\right]\nonumber \\
&\times&\left[(t^{a}S(x))_{ij}(S(y)t^{b})_{kl}+(S(x)t^{b})_{ij}(t^{a}S(y))_{kl}\right]\,.
\end{eqnarray}

Under  $z$ and $z^\prime$ integrals, and using various relations between the kernels (\ref{Krelate}), all the terms  can be equivalently written as:
\begin{eqnarray}
{\cal A}_1&=& {\cal A}_{10}+{\cal A}_{11}+{\cal A}_{12};\nonumber \\ 
&&{\cal A}_{10}=\left[(t^{a}S(x))_{ij}(S(y)t^{b})_{kl}+(S(x)t^{b})_{ij}(t^{a}S(y))_{kl}\right] 
\nonumber \\
&&\times\Big\{ - \bar K_{JSSJ}(x,y,z,z^{\prime}) S_{A}^{dd^{\prime}}(z)\left[S_{A}^{ee^{\prime}}(z^{\prime})-S_{A}^{ee^{\prime}}(z)\right] f^{ade}f^{bd^{\prime}e^{\prime}}\nonumber \\
&&-2\bar K_{q\bar{q}}(x,y,z,z^{\prime}) tr\left[t^{a}S(z)t^{b}(S(z^{\prime})^{\dagger}-S(z)^{\dagger})\right]\Big\};\nonumber \\
&&{\cal A}_{11}=S_{A}^{dd^{\prime}}(z)S_{A}^{ee^{\prime}}(z^{\prime})
 \Big\{ \left[f^{ad^{\prime}e^{\prime}}(S(x)t^{a})_{ij}(t^{d}t^{e}S(y))_{kl}-f^{ade}(t^{a}S(x))_{ij}(S(y)t^{e^{\prime}}t^{d^{\prime}})_{kl}\right] \nonumber \\
&&\times \left( K_{JJSSJ}(y,y,x,z,z^{\prime})-K_{JJSSJ}(y,x,y,z,z^{\prime})\right)\nonumber \\
&&+ \left[f^{ad^{\prime}e^{\prime}}(t^{d}t^{e}S(x))_{ij}(S(y)t^{a})_{kl}-f^{ade}(S(x)t^{e^{\prime}}t^{d^{\prime}})_{ij}(t^{a}S(y))_{kl}\right]\nonumber \\
&&\times \left( K_{JJSSJ}(x,x,y,z,z^{\prime})-K_{JJSSJ}(x,y,x,z,z^{\prime})\right)\Big\};\nonumber \\
&&{\cal A}_{12}=- N_c\,S_A^{ab}(z) \left[(t^{a}S(x))_{ij}(S(y)t^{b})_{kl}+(S(x)t^{b})_{ij}(t^{a}S(y))_{kl}\right]  \widetilde{K}(x,y,z,z^{\prime}).
\end{eqnarray}
\begin{eqnarray}
{\cal A}_2&=& {\cal A}_{20}+{\cal A}_{21}+{\cal A}_{22}+{\cal A}_{23}; \nonumber \\ 
&&{\cal A}_{20}=- 2\,S_{A}^{dd^{\prime}}(z) S_{A}^{ee^{\prime}}(z^{\prime})\nonumber \\
&&
\times \Big\{\left[f^{ad^{\prime}e^{\prime}}(t^{d}S(x)t^{a})_{ij}(t^{e}S(y))_{kl}-f^{ade}(t^{a}S(x)t^{d^{\prime}})_{ij}(S(y)t^{e^{\prime}})_{kl}\right]K_{JJSSJ}(x,x,y,z,z^{\prime})\nonumber \\
&&+\left[f^{ad^{\prime}e^{\prime}}(t^{d}S(x))_{ij}(t^{e}S(y)t^{a})_{kl}-f^{ade}(S(x)t^{d^{\prime}})_{ij}(t^{a}S(y)t^{e^{\prime}})_{kl}\right]K_{JJSSJ}(y,x,y,z,z^{\prime})\Big\}; \nonumber \\
&&{\cal A}_{21}= 2\,S_{A}^{dd^{\prime}}(z) \nonumber \\
&&\times \Big\{\left[f^{ad^{\prime}e^\prime}(t^{d}S(x)t^{a})_{ij}(S(y)t^{e^\prime})_{kl}-f^{ade}(t^{a}S(x)t^{d^{\prime}})_{ij}(t^{e}S(y))_{kl}\right]K_{JJSSJ}(x,x,y,z,z^{\prime})\nonumber \\
&&- \left[f^{ae^{\prime}d^{\prime}}(S(x)t^{e^\prime})_{ij}(t^{d}S(y)t^{a})_{kl}-f^{aed}(t^{e}S(x))_{ij}(t^{a}S(y)t^{d^{\prime}})_{kl}\right] K_{JJSSJ}(y,y,x,z,z^{\prime})\Big\}; \nonumber \\
&&{\cal A}_{22}={iN_c\over 2} \left[ (t^{d}S(x))_{ij}(S(y)t^{d^{\prime}})_{kl}+(S(x)t^{d^{\prime}})_{ij}(t^{d}S(y))_{kl}\right]\nonumber \\
&&\times (K_{JJSSJ}(y,x,y,z,z^{\prime})+ K_{JJSSJ}(x,y,x,z,z^{\prime}))=0;\nonumber \\
&&{\cal A}_{23}=-2  \Big\{\left[f^{ad^{\prime}e^{\prime}}(S(x)t^{d^{\prime}}t^{a})_{ij}(S(y)t^{e^\prime})_{kl}-f^{ade}(t^{a}t^{d}S(x))_{ij}(t^{e}S(y))_{kl}\right]\nonumber \\
&&\times K_{JJSSJ}(x,x,y,z,z^{\prime})
+\left[f^{ad^{\prime}e^{\prime}}(S(x)t^{d^\prime})_{ij}(S(y)t^{e^\prime}t^{a})_{kl}-f^{ade}(t^{d}S(x))_{ij}(t^{a}t^{e}S(y))_{kl}\right]\nonumber \\
&&\times K_{JJSSJ}(y,x,y,z,z^{\prime})\Big\}=0 .
\end{eqnarray}
\begin{eqnarray}
{\cal A}_3&=&{\cal A}_{30} +{\cal A}_{31}+{\cal A}_{32}; \nonumber \\ 
&&{\cal A}_{30}= -S_{A}^{dd^{\prime}}(z) S_{A}^{ee^{\prime}}(z^{\prime})
\Big\{ \left[f^{ad^{\prime}e^{\prime}}(S(x)t^{a})_{ij}(t^{d}t^{e}S(y))_{kl}-f^{ade}(t^{a}S(x))_{ij}(S(y)t^{e^{\prime}}t^{d^{\prime}})_{kl}\right]
 \nonumber \\ 
&&\times  K_{JJSSJ}(x,y,y,z,z^{\prime})\nonumber \\
&&+ \left[f^{ad^{\prime}e^{\prime}}(t^{d}t^{e}S(x))_{ij}(S(y)t^{a})_{kl}-f^{ade}(S(x)t^{e^{\prime}}t^{d^{\prime}})_{ij}(t^{a}S(y))_{kl}\right] K_{JJSSJ}(y,x,x,z,z^{\prime})\Big\}; \nonumber \\
&&{\cal A}_{31}=-{\cal A}_{11}; \nonumber \\
&&{\cal A}_{32}= S_{A}^{dd^{\prime}}(z)\Big\{ \left[f^{ad^{\prime}e^{\prime}}(S(x)t^{a})_{ij}(t^{d}S(y)t^{e^\prime})_{kl}-f^{ade}(t^{a}S(x))_{ij}(t^{e}S(y)t^{d^{\prime}})_{kl}\right]\nonumber \\
&&\times \left( K_{JJSSJ}(x,y,y,z,z^{\prime})+K_{JJSSJ}(y,y,x,z,z^{\prime})-K_{JJSSJ}(y,x,y,z,z^{\prime})\right)\nonumber \\
&&+ \left[f^{ad^{\prime}e^{\prime}}(t^{d}S(x)t^{e\prime})_{ij}(S(y)t^{a})_{kl}-f^{ade}(t^{e}S(x)t^{d^{\prime}})_{ij}(t^{a}S(y))_{kl}\right]\ \nonumber \\
&&\times\left( K_{JJSSJ}(y,x,x,z,z^{\prime})+K_{JJSSJ}(x,x,y,z,z^{\prime})-K_{JJSSJ}(x,y,x,z,z^{\prime})\right)\Big\}.
\end{eqnarray}
\begin{eqnarray}
{\cal B}_1&=& {\cal B}_{10}+{\cal B}_{11}; \nonumber \\ 
&&{\cal B}_{10}=-S_{A}^{ab}(z) \Big\{\left[f^{bde}(t^{a}S(x)t^{d})_{ij}(S(y)t^{e})_{kl}-f^{ade}(t^{d}S(x)t^{b})_{ij}(t^{e}S(y))_{kl}\right] \nonumber \\
&&\times K_{JJSJ}(x,y,x,z)\nonumber \\
&&+ \left[f^{bde}(S(x)t^{e})_{ij}(t^{a}S(y)t^{d})_{kl}-f^{ade}(t^{e}S(x))_{ij}(t^{d}S(y)t^{b})_{kl}\right]K_{JJSJ}(y,x,y,z)
\Big\} ; \nonumber \\
&&{\cal B}_{11}=\left[f^{bde}(S(x)t^{b}t^{d})_{ij}(S(y)t^{e})_{kl}-f^{ade}(t^{d}t^{a}S(x))_{ij}(t^{e}S(y))_{kl}\right] K_{JJSJ}(x,y,x,z) \nonumber \\
&&+ \left[f^{bde}(S(x)t^{e})_{ij}(S(y)t^{b}t^{d})_{kl}-f^{ade}(t^{e}S(x))_{ij}(t^{d}t^{a}S(y))_{kl}\right]K_{JJSJ}(y,x,y,z)=0.\nonumber \\
\end{eqnarray}
\begin{eqnarray}
{\cal B}_2&=& {\cal B}_{20}+{\cal B}_{21}+{\cal B}_{22};\nonumber \\ 
&&{\cal B}_{20}=- \bar K_{JSJ}(x,y,z)\left[S_{A}^{ab}(x)+S_{A}^{ab}(y)-2S_{A}^{ab}(z)\right] \nonumber \\
&&\times \left[(t^{a}S(x))_{ij}(S(y)t^{b})_{kl}+(S(x)t^{b})_{ij}(t^{a}S(y))_{kl}\right];\nonumber \\
&&{\cal B}_{21}=-{\cal A}_{12};\nonumber \\
&&{\cal B}_{22}= -N_c \int_{z^{\prime}}\widetilde{K}(x,y,z,z^{\prime}) 
\left[(S(x)t^{a})_{ij}(S(y)t^{a})_{kl}+(t^{a}S(x))_{ij}(t^{a}S(y))_{kl}\right]=0.\nonumber \\
\end{eqnarray}

Acting  on the two Wilson lines with our Hamiltonian:
\begin{eqnarray}
\frac{d}{dY}(S(x))_{ij}(S(y))_{kl}\mid_{conn.}&=&-H^{NLO\ JIMWLK}(S(x))_{ij}(S(y))_{kl}\mid_{conn.}\nonumber \\
&=&{\cal A}_{10}+{\cal A}_{20}+{\cal A}_{30}+2 {\cal B}_{10}+{\cal B}_{20}\,.
\end{eqnarray}
We complete the proof of  equivalence by noticing that ${\cal A}_{21}+{\cal A}_{32}={\cal B}_{10}$.

 \subsection{Triple interaction}

The connected part of three-Wilson line evolution according to \cite{BClast} is
\begin{equation}\begin{split}\label{triple}
&\frac{d}{dY}(S(x))_{ij}(S(y))_{kl}(S(w))_{mn}\mid_{conn.}=-2\int_{z,z^{\prime}}K_{JJSSJ}(w,x,y;z,z^{\prime})\left[S_{A}^{ad}(z)S_{A}^{be}(z^{\prime})\right. \\
&\left. -S_{A}^{ad}(x)S_{A}^{be}(z^{\prime})-S_{A}^{ad}(z)S_{A}^{be}(y)+S_{A}^{ad}(x)S_{A}^{be}(y)\right] \left[f^{cde}(t^{a}S(x))_{ij}(t^{b}S(y))_{kl}(S(w)t^{c})_{mn}\right. \\
&\left. -f^{cab}(S(x)t^{d})_{ij}(S(y)t^{e})_{kl}(t^{c}S(w))_{mn}\right]+\left[w,(mn)\leftrightarrow y,(kl)\right]+\left[w,(mn)\leftrightarrow x,(ij)\right]=\\
&=-2\int_{z,z^{\prime}}K_{JJSSJ}(w,x,y;z,z^{\prime})S_{A}^{ad}(z)S_{A}^{be}(z^{\prime})\left[f^{cde}(t^{a}S(x))_{ij}(t^{b}S(y))_{kl}(S(w)t^{c})_{mn}\right. \\
&\left.-f^{cab}(S(x)t^{d})_{ij}(S(y)t^{e})_{kl}(t^{c}S(w))_{mn}\right]+2\int_{z}K_{JJSJ}(x,y,w,z)S_{A}^{ad}(z)\\
&\times \left[f^{cde}(t^{a}S(x))_{ij}(S(y)t^{e})_{kl}(S(w)t^{c})_{mn}-f^{cab}(S(x)t^{d})_{ij}(t^{b}S(y))_{kl}(t^{c}S(w))_{mn}\right]\\
&-2\int_{z,z^{\prime}}K_{JJSSJ}(w,x,y;z,z^{\prime})f^{cde}\left[(S(x)t^{d})_{ij}(S(y)t^{e})_{kl}(S(w)t^{c})_{mn}\right. \\
&\left. -(t^{d}S(x))_{ij}(t^{e}S(y))_{kl}(t^{c}S(w))_{mn}\right]
+\left[w,(mn)\leftrightarrow y,(kl)\right]+\left[w,(mn)\leftrightarrow x,(ij)\right]\,.
\end{split}\end{equation}
Here the second equality was obtained using (\ref{Krelate}).
In the NLO  Hamiltonian, the connected part of the evolution of three $S$ originates from the terms containing three $J$s only, when each $J$ acts 
on a different $S$. So, we now retain only these terms. From the action of the Hamiltonian we find:
 \begin{equation}\begin{split}
&\frac{d}{dY}(S(x))_{ij}(S(y))_{kl}(S(w))_{mn}\mid_{conn.}=-H^{NLO\ JIMWLK}(S(x))_{ij}(S(y))_{kl}(S(w))_{mn}\mid_{conn.}=\\
&=-\Big\{\int_{z,z^{\prime}}K_{JJSSJ}(w,x,y;z,z^{\prime})S_{A}^{dc}(z)S_{A}^{eb}(z^{\prime})\left[f^{acb}(t^{d}S(x))_{ij}(t^{e}S(y))_{kl}(S(w)t^{a})_{mn}\right.\\
&\left.-f^{ade}(S(x)t^{c})_{ij}(S(y)t^{b})_{kl}(t^{a}S(w))_{mn}\right]
+\int_{z}K_{JJSJ}(x,y,w;z)S_{A}^{ba}(z) \\
&\times  \left[f^{bde}(S(x)t^{a})_{ij}(t^{d}S(y))_{kl}(t^{e}S(w))_{mn}-f^{ade}(t^{b}S(x))_{ij}(S(y)t^{d})_{kl}(S(w)t^{e})_{mn}\right] \\
&+{1\over 3}\int_{z,z^{\prime}}\left(K_{JJSSJ}(w,x,y;z,z^{\prime})+K_{JJSSJ}(y,w,x;z,z^{\prime})-K_{JJSSJ}(x,w,y;z,z^{\prime})\right)\\ 
& \times f^{acb}\left[(S(x)t^{c})_{ij}(S(y)t^{b})_{kl}(S(w)t^{a})_{mn}
 -(t^{c}S(x))_{ij}(t^{b}S(y))_{kl}(t^{a}S(w))_{mn}\right]\\
 &+\left[x,(ij)\leftrightarrow y,(kl)\right]\Big\}+\left[w,(mn)\leftrightarrow y,(kl)\right]+\left[w,(mn)\leftrightarrow x,(ij)\right]\,.
 \end{split}\end{equation}
Here we have used the relation (\ref{Krelate}). Using the symmetry properties of the kernels, we can rewrite this as (\ref{triple})
thus reproducing the result of \cite{BClast}.
This concludes our comparison and establishes a complete equivalence between the approach based on the NLO JIMWLK 
Hamiltonian and a set of evolution equations for the Wilson lines.

\section{NLO evolution of three-quark Wilson loop}

In this Section we derive the evolution equation for the three-quark operator $B$ defined in the Introduction.  As a warm up we reproduce the Leading Order result of \cite{GrabLO}, and then consider the Next to Leading Order.  As explained above, some of the NLO results were computed in \cite{Grab} and are used as input to determine
the kernels $K_{JJSSJ}$ and $K_{JJSJ}$.  The rest of the results of this Section are new. We derive complete NLO evolution of $B$ in QCD
and also the evolution of its conformal extension $\cal B$ in ${\cal N}=4 $ theory.

We note that although we refer to $B$ as a three quark operator, or baryon, it in fact can be expressed in terms of a quadrupole operator with two coinciding points. This is due to the fact that one of the quark Wilson lines in $SU(3)$ theory can be written as a product of two antifundamental Wilson lines
\begin{equation}
S_{ij}(x)=\frac{1}{2}\epsilon_{ijk}\epsilon_{jmn}S^\dagger(x)_{mk}S^\dagger(x)_{nl}\,.
\end{equation}
Using this $SU(3)$ identity we find
\beq\label{quad}
B(x,y,z)=tr[S(x)S^\dagger(z)]tr[S(y)S^\dagger(z)]-tr[S(x)S^\dagger(z)S(y)S^\dagger(z)]\,.
\eeq

\subsection{LO evolution}

The LO evolution of $B$ is driven by the equation 
\beq
\frac{d}{dY}\,B=-H^{LO\,JIMWLK}\,B\,.
\eeq
Due to relation (\ref{quad}), the Leading Order evolution of $B$ can be read off the known evolution of the quadrupole operator \cite{quadrupole1,quadrupole2}. 
Nevertheless, for the sake of completeness we re-derive this equation here.

We start from computing the action of the left and right rotation generators on $B$:
\begin{equation}\begin{split}
&J_{L}^{a}(w)B(x,y,z)=\varepsilon^{ijk}\varepsilon^{lmn}\left[(t^{a}S(x))^{il}S^{jm}(y)S^{kn}(z)\delta(x-w)\right. \\
&\ \ \ \ \ \ \left.+S^{il}(x)(t^{a}S(y))^{jm}S^{kn}(z)\delta(y-w)
 +S^{il}(x)S^{jm}(y)(t^{a}S(z))^{kn}\delta(z-w)\right], \\
&J_{R}^{a}(w)B(x,y,z)=\varepsilon^{ijk}\varepsilon^{lmn}\left[(S(x)t^{a})^{il}S^{jm}(y)S^{kn}(z)\delta(x-w)\right. \\
&\ \ \ \ \ \ \ \left.+S^{il}(x)(S(y)t^{a})^{jm}S^{kn}(z)\delta(y-w)+S^{il}(x)S^{jm}(y)(S(z)t^{a})^{kn}\delta(z-w)\right],\\
\end{split}\end{equation}
and
\begin{equation}\begin{split}
&J_{L}^{a}(v)J_{L}^{a}(w)B(x,y,z)=\varepsilon^{ijk}\varepsilon^{lmn}\delta(x-w)\left[(t^{a}t^{a}S(x))^{il}S^{jm}(y)S^{kn}(z)\delta(x-v)\right.\\
&\left.+(t^{a}S(x))^{il}(t^{a}S(y))^{jm}S^{kn}(z)\delta(y-v)(t^{a}S(x))^{il}S^{jm}(y)(t^{a}S(z))^{kn}\delta(z-v)\right]\\
&+\varepsilon^{ijk}\varepsilon^{lmn}\delta(y-w)\left[(t^{a}S(x))^{il}(t^{a}S(y))^{jm}S^{kn}(z)\delta(x-v)\right.\\
&+\left.S^{il}(x)(t^{a}t^{a}S(y))^{jm}S^{kn}(z)\delta(y-v)+S^{il}(x)(t^{a}S(y))^{jm}(t^{a}S(z))^{kn}\delta(z-v)\right]\\
&+\varepsilon^{ijk}\varepsilon^{lmn}\delta(z-w)\left[(t^{a}S(x))^{il}S^{jm}(y)(t^{a}S(z))^{kn}\delta(x-v)\right.\\
&+\left.S^{il}(x)(t^{a}S(y))^{jm}(t^{a}S(z))^{kn}\delta(y-v)+S^{il}(x)S^{jm}(y)(t^{a}t^{a}S(z))^{kn}\delta(z-v)\right]\\
&=\frac{N_{c}+1}{2N_{c}}B(x,y,z)\Big(\left[(N_{c}-1)\delta(x-v)-\delta(y-v)-\delta(z-v)\right]\delta(x-w)\Big)\\ 
&+\left(x\leftrightarrow y\right)+\left(x\leftrightarrow z\right)\,.\\
\end{split}\end{equation}
Similarly:
\begin{equation}\begin{split}
&J_{R}^{a}(v)J_{R}^{a}(w)B(x,y,z)
=\frac{N_{c}+1}{2N_{c}}B(x,y,z)\\
&\ \ \ \ \ \ \ \ \times \left(\left[(N_{c}-1)\delta(x-v)-\delta(y-v)-\delta(z-v)\right]\delta(x-w)+\left(x\leftrightarrow y\right)+\left(x\leftrightarrow z\right)\right)\,.\\
\end{split}\end{equation}
The contribution from the real term yields
\begin{equation}\begin{split}
&2S_{A}^{ab}(u)J_{L}^{a}(v)J_{R}^{b}(w)B(x,y,z)=\\
&4\varepsilon^{ijk}\varepsilon^{lmn}\delta(x-w)\left[(t^{a}S(x)t^{b})^{il}S^{jm}(y)S^{kn}(z)\delta(x-v)+(S(x)t^{b})^{il}(t^{a}S(y))^{jm}S^{kn}(z)\delta(y-v)\right.\\
&\left.+(S(x)t^{b})^{il}S^{jm}(y)(t^{a}S(z))^{kn}\delta(z-v)\right]\times tr\left[t^{a}S(u)t^{b}S^{\dagger}(u)\right]+\left(x\leftrightarrow y\right)+\left(x\leftrightarrow z\right)=\\
&=\delta(x-w)\left[\left(N_{c}s(u,x)B(u,y,z)-\frac{1}{N_{c}}B(x,y,z)\right)\delta(x-v)\right. \\
&\left. +\left(-(S(x)S^{\dagger}(u)S(y))\cdot S(u)\cdot S(z)-\frac{1}{N_{c}}B(x,y,z)\right)\delta(y-v)\right.\\
&\left.+\left(-(S(x)S^{\dagger}(u)S(z))\cdot S(y)\cdot S(u)-\frac{1}{N_{c}}B(x,y,z)\right)\delta(z-v)\right]+(x\leftrightarrow y)+(x\leftrightarrow z)\,.\\
\end{split}\end{equation}
Here, following \cite{Grab}, we have utilized the notation:
\begin{equation}\begin{split}
S(x)\cdot S(y)\cdot S(z)=\varepsilon^{ikm}\varepsilon^{jln}S^{ij}(x)S^{kl}(y)S^{mn}(z)\,.
 \end{split}\end{equation}
All together we have:
\begin{equation}\begin{split}\label{LOB}
&-H^{LO\, JIMWLK}\,B(x,y,z)= \\
&={1\over 2}\int_{v,w,u}M(v,w,u)\left[J_{L}^{a}(v)J_{L}^{a}(w)+J_{R}^{a}(v)J_{R}^{a}(w)-2J_{L}^{a}(v)S_A^{ab}(u)J_{R}^{b}(w)\right]B(x,y,z)=\\
&={1\over 2}\int_{u}M(x,y,u)\Big[(S(x)S^{\dagger}(u)S(y))\cdot S(u)\cdot S(z)+(S(y)S^{\dagger}(u)S(x))\cdot S(u)\cdot S(z) \\
& \ \ \ \ \ \ \  \ \ \ \ \ -2B(x,y,z)\Big]+(x\leftrightarrow z) +(y\leftrightarrow z)\,.
  \end{split}\end{equation}
The following identity \cite{Grab} holds:
\begin{equation}\begin{split}\label{Grabidentity}
&(S(x)S^{\dagger}(u)S(y))\cdot S(u)\cdot S(z)+(S(y)S^{\dagger}(u)S(x))\cdot S(u)\cdot S(z)\equiv \\
&-B(x,y,z)+\frac{1}{2}\Big(B(x,u,u)B(z,y,u)+B(y,u,u)B(z,x,u)-B(z,u,u)B(y,x,u)\Big)\,.\\
\end{split}\end{equation}
Inserting this identity into (\ref{LOB}) yields:
\begin{equation}\begin{split}
&{d\over dY} B(x,y,z)={3\over 2} \int_{u}M(x,y,u)\Big[-B(x,y,z)+\frac{1}{6}\left(B(x,u,u)B(z,y,u)\right.\\ 
&\left.\ \ \ \ \ \ \ \ \ \ \ \ \ \ \ \ \ \ \ \ \ \ \ \ +B(y,u,u)B(z,x,u)-B(z,u,u)B(y,x,u)\right)\Big]+(x\leftrightarrow z) +(y\leftrightarrow z)\,.\\
 \end{split}\end{equation}
In this form the equation appears in \cite{Grab,GrabLO}.  Note, that it can be simplified using
\beq
B(x,u,u)\,=\,6\,s(x,u)\ .
\eeq
Hence
\begin{equation}\begin{split}
&{d\over dY} B(x,y,z)={3\over 2} \int_{u}M(x,y,u)\Big[-B(x,y,z)+s(x,u)B(z,y,u)\\ 
&\ \ \ \ \ \ \ \ \ \ \ \ \ \ \ \ \ \ \ \ \ \ \ \ + s(y,u)B(z,x,u)-s(z,u)B(y,x,u)\Big]+(x\leftrightarrow z) +(y\leftrightarrow z)\,.\\
 \end{split}\end{equation}
A linearized version of this equation was also derived in \cite{Hattaoderon}.
  
\subsection{NLO evolution}
  At NLO the evolution of $B$ is given by 
\beq\label{NLOB}
\frac{d}{dY}\,B=-H^{NLO\,JIMWLK}\,B= -\left[\dot B_{JSJ} +\dot B_{JSSJ}+\dot B_{JJSJ}+\dot B_{JJSSJ}+\dot B_{qq}\right]\,,
\eeq
where the notation is self explanatory.
All the terms that appear in the rhs of (\ref{NLOB}) are computed below.

 \subsection*{$\bf K_{JJSJ}$}
 
We start with the $K_{JJSJ}$ term in the Hamiltonian. When three $J$s in the Hamiltonian act on three $S$ in $B$ they  produce fully connected terms, that
is terms where each $J$ acts on a different $S$,  and virtual terms when at least one of the $S$ is left untouched. The former  terms were computed in \cite{Grab} 
and were used by us as an input to fix the kernel $K_{JJSJ}$. The virtual terms calculated below are new.

 The action on $B$ reads  
 \begin{equation}\begin{split}
&f^{bde}J_{L}^{d}(v)J_{L}^{e}(u)S_{A}^{ba}(p)J_{R}^{a}(w)B(x,y,z)=f^{bde}S_{A}^{ba}(p)\varepsilon^{ijk}\varepsilon^{lmn}\times\\
&\left(\left[(t^{e}t^{d}S(x)t^{a})^{il}S^{jm}(y)S^{kn}(z)\delta(x-v)+(t^{e}S(x)t^{a})^{il}(t^{d}S(y))^{jm}S^{kn}(z)\delta(y-v)+\right.\right.\\
&\left.+(t^{e}S(x)t^{a})^{il}S^{jm}(y)(t^{d}S(z))^{kn}\delta(z-v)\right]\delta(x-u)\delta(x-w)+\\
&\left[(t^{d}S(x)t^{a})^{il}(t^{e}S(y))^{jm}S^{kn}(z)\delta(x-v)+(S(x)t^{a})^{il}(t^{e}t^{d}S(y))^{jm}S^{kn}(z)\delta(y-v)+\right.\\
&\left.+(S(x)t^{a})^{il}(t^{e}S(y))^{jm}(t^{d}S(z))^{kn}\delta(z-v)\right]\delta(y-u)\delta(x-w)+\\
&\left[(t^{\ d}S(x)t^{a})^{il}S^{jm}(y)(t^{e}S(z))^{kn}\delta(x-v)+(S(x)t^{a})^{il}(t^{d}S(y))^{jm}(t^{e}S(z))^{kn}\delta(y-v)+\right.\\
&\left.\left.+(S(x)t^{a})^{il}S^{jm}(y)(t^{e}t^{d}S(z))^{kn}\delta(z-v)\right]\delta(z-u)\delta(x-w)\right)+\left(x\leftrightarrow y\right)+\left(x\leftrightarrow z\right).\\
\end{split}\end{equation}
Similarly, the $LRR$ term:
 \begin{equation}\begin{split}
&f^{bde}J_{L}^{a}(w)S_{A}^{ab}(p)J_{R}^{d}(v)J_{R}^{e}(u)B(x,y,z)=f^{ade}S_{A}^{ba}(p)\varepsilon^{ijk}\varepsilon^{lmn}\times\\
&\left(\left[(t^{b}S(x)t^{d}t^{e})^{il}S^{jm}(y)S^{kn}(z)\delta(x-v)+(t^{b}S(x)t^{e})^{il}(S(y)t^{d})^{jm}S^{kn}(z)\delta(y-v)+\right.\right.\\
&\left.+(t^{b}S(x)t^{e})^{il}S^{jm}(y)(S(z)t^{d})^{kn}\delta(z-v)\right]\delta(x-u)\delta(x-w)+\\
&\left[(t^{b}S(x)t^{d})^{il}(S(y)t^{e})^{jm}S^{kn}(z)\delta(x-v)+(t^{b}S(x))^{il}(S(y)t^{d}t^{e})^{jm}S^{kn}(z)\delta(y-v)+\right.\\
&\left.+(t^{b}S(x))^{il}(S(y)t^{e})^{jm}(S(z)t^{d})^{kn}\delta(z-v)\right]\delta(y-u)\delta(x-w)+\\
&\left[(t^{b}S(x)t^{d})^{il}S^{jm}(y)(S(z)t^{e})^{kn}\delta(x-v)+(t^{b}S(x))^{il}(S(y)t^{d})^{jm}(S(z)t^{e})^{kn}\delta(y-v)+\right.\\
&\left.\left.+(t^{b}S(x))^{il}S^{jm}(y)(S(z)t^{d}t^{e})^{kn}\delta(z-v)\right]\delta(z-u)\delta(x-w)\right)+\left(x\leftrightarrow y\right)+\left(x\leftrightarrow z\right). \\
\end{split}\end{equation}
We should now multiply by the kernel $K_{JJSJ}$ and perform the integrations. First let us zoom in on the connected terms:
\begin{equation}\begin{split}
&\dot B_{JJSJ}^{connected}\equiv\int_{w,v,u,p}K_{JJSJ}(w;v,u;p)f^{bde}\left[J_{L}^{d}(v)J_{L}^{e}(u)S_{A}^{ba}(p)J_{R}^{a}(w)\right. \\
&\left. \ \ \ \ \ \ \ \ \ \ \ \ \ \ \ \ \ \ \ \ \ -\ J_{L}^{a}(w)S_{A}^{ab}(p)J_{R}^{d}(v)J_{R}^{e}(u)\right]B(x,y,z)\mid_{conn.}\\
&=2\int_{p}K_{JJSJ}(x;y,z;p) S_{A}^{ba}(p)\varepsilon^{ijk}\varepsilon^{lmn}\Big[f^{bde}(S(x)T^{a})^{il}(T^{d}S(y))^{jm}(T^{e}S(z))^{kn}\\
&-f^{ade}(T^{b}S(x))^{il}(S(y)T^{d})^{jm}(S(z)T^{e})^{kn}Big]+\left(x\leftrightarrow y\right)+\left(x\leftrightarrow z\right)\\
&=i {1\over 2} \int_{p}K_{JJSJ}(x;y,z;p)\Big[(S(x)S^{\dagger}(p)S(z))\cdot S(p)\cdot S(y)-S(x)\cdot(S(y)S^{\dagger}(p)S(z))\cdot S(p) \\
&+(S(z)S^{\dagger}(p)S(x))\cdot S(p)\cdot S(y)-S(x)\cdot(S(z)S^{\dagger}(p)S(y))\cdot S(p)
\Big]+\left(x\leftrightarrow y\right)+\left(x\leftrightarrow z\right)\\
&={-3i} \int_{p}K_{JJSJ}(x;y,z;p)\Big[s(y,p)B(z,x,p)-s(z,p)B(x,y,p)\Big]+\left(x\leftrightarrow y\right)+\left(x\leftrightarrow z\right)\,.
 \end{split}\end{equation}
Comparing this with eq. (5.25) of \cite{Grab} we read off the kernel $K_{JJSJ}$ as quoted in (\ref{KJJSJ}).

The contribution of the disconnected part is:
 \begin{equation}\begin{split}
&\dot B_{JJSJ}^{disconnected}\equiv
\int_{w,v,u,p}K_{JJSJ}(w;v,u;p)f^{bde}\left[J_{L}^{d}(v)J_{L}^{e}(u)S_{A}^{ba}(p)J_{R}^{a}(w)\right. \\
&\ \ \ \ \ \ \ \ \ \ \ \ \ \ \ \  \ \ \  \ \ \ \ \ \ \ \left. -J_{L}^{a}(w)S_{A}^{ab}(p)J_{R}^{d}(v)J_{R}^{e}(u)\right]B(x,y,z)\mid_{disconn.}=\\
&=\frac{i}{2}\int_{p} \Big\{ K_{JJSJ}(x;y,x;p)\left[ 6 s(x,p) B(y,z,p)+ S(p)\cdot(S(x)S^{\dagger}(p)S(y))\cdot S(z)\right. \\
& \left. +S(p)\cdot(S(y)S^{\dagger}(p)S(x))\cdot S(z)  \right]+K_{JJSJ}(x;z,x;p)\left[S(p)\cdot(S(x)S^{\dagger}(p)S(z))\cdot S(y)\right. \\
&\left.  +S(p)\cdot(S(z)S^{\dagger}(p)S(x))\cdot S(y)+6 s(x,p) B(y,z,p)  \right]\Big\}+\left(x\leftrightarrow y\right)+\left(x\leftrightarrow z\right)\\
&=-\frac{i}{2}\int_{p} \Big\{ K_{JJSJ}(x;y,x;p)[ B(x,y,z) +9s(x,p) B(y,z,p)- 3 s(y,p) B(x,z,p) \\ 
&+ 3s(z,p) B(x,y,p)] +K_{JJSJ}(x;z,x;p)[ B(x,y,z) +9s(x,p) B(y,z,p)- 3 s(z,p) B(x,y,p) \\ 
&+ 3s(y,p) B(x,z,p)]
 \Big\}+\left(x\leftrightarrow y\right)+\left(x\leftrightarrow z\right)\,.\\
 \end{split}\end{equation}
We have  re-expressed the result in terms of the operator $B$ using the identity (\ref{Grabidentity}). In this final form the disconnected
part is found to agree with \cite{Grab}.

Next up are the virtual terms:
\begin{equation}\begin{split}
&f^{bde}J_{L}^{d}(v)J_{L}^{e}(u)J_{L}^{b}(w)B(x,y,z)=f^{bde}\varepsilon^{ijk}\varepsilon^{lmn}\left(\left[(t^{b}t^{e}t^{d}S(x))^{il}S^{jm}(y)S^{kn}(z)\delta(x-v)\right.\right.\\
&\left. +(t^{b}t^{e}S(x))^{il}(t^{d}S(y))^{jm}S^{kn}(z)\delta(y-v)+(t^{b}t^{e}S(x))^{il}S^{jm}(y)(t^{d}S(z))^{kn}\delta(z-v)\right]\\ 
&\times \delta(x-u)\delta(x-w)+
\left[(t^{b}t^{d}S(x))^{il}(t^{e}S(y))^{jm}S^{kn}(z)\delta(x-v)+(t^{b}S(x))^{il}(t^{e}t^{d}S(y))^{jm}\right. \\
&\times S^{kn}(z)\delta(y-v) \left.+(t^{b}S(x))^{il}(t^{e}S(y))^{jm}(t^{d}S(z))^{kn}\delta(z-v)\right]\delta(y-u)\delta(x-w)\\
&+\left[(t^{b}t^{d}S(x))^{il}S^{jm}(y)(t^{e}S(z))^{kn}\delta(x-v)+(t^{b}S(x))^{il}(t^{d}S(y))^{jm}(t^{e}S(z))^{kn}\delta(y-v)\right.\\
&\left.\left.+(t^{b}S(x))^{il}S^{jm}(y)(t^{e}t^{d}S(z))^{kn}\delta(z-v)\right]\delta(z-u)\delta(x-w)\right)+\left(x\leftrightarrow y\right)+\left(x\leftrightarrow z\right)=\\
&=\frac{i}{4}B(x,y,z)\left(\left[-(N_{c}^{2}-1)\delta(x-v)+(N_{c}+1)\delta(y-v)+(N_{c}+1)\delta(z-v)\right]\delta(x-u)\right. \\
&\times \delta(x-w)
+\left[(N_{c}+1)\delta(x-v)+(N_{c}+1)\delta(y-v)\right]\delta(y-u)\delta(x-w)\\
&+\left.\left[(N_{c}+1)\delta(x-v)+(N_{c}+1)\delta(z-v)\right]\delta(z-u)\delta(x-w)\right)+\left(x\leftrightarrow y\right)+\left(x\leftrightarrow z\right)\,.\\
\end{split}\end{equation}
Multiplying by the kernel $K_{JJSJ}$ and integrating we get
\begin{equation}\begin{split}
&\int_{w,v,u,p}K_{JJSJ}(w,v,u,p)f^{bde}J_{L}^{d}(v)J_{L}^{e}(u)J_{L}^{b}(w)B(x,y,z)=\\
&=\int_{p}\frac{i}{4}(N_{c}+1)\left[K_{JJSJ}(x,y,x,p)+K_{JJSJ}(x,z,x,p)+K_{JJSJ}(x,x,y,p)+K_{JJSJ}(x,x,z,p)\right.\\
&\left.+\left(x\leftrightarrow y\right)+\left(x\leftrightarrow z\right)\right]=0\,.
\end{split}\end{equation}
We find that there is no contribution to the evolution of $B$ from the three-$J$ virtual term, as noted in the introduction.
Thus
\beq
\dot B_{JJSJ}=\dot B_{JJSJ}^{connected}+\dot B_{JJSJ}^{disconnected}\,.
\eeq

\subsection*{$\bf K_{JJSSJ}$}

Continuing with the kernel $K_{JJSSJ}$, we obtain: 
\begin{equation}\begin{split}
&f^{acb}J_{L}^{d}(r)J_{L}^{e}(v)S_{A}^{dc}(u)S_{A}^{eb}(u^{\prime})J_{R}^{a}(w)B(x,y,z)=f^{acb}S_{A}^{dc}(u)S_{A}^{eb}(u^{\prime})\varepsilon^{ijk}\varepsilon^{lmn}\\
&\times \left(\left[(t^{e}t^{d}S(x)t^{a})^{il}S^{jm}(y)S^{kn}(z)\delta(x-r)+(t^{e}S(x)t^{a})^{il}(t^{d}S(y))^{jm}S^{kn}(z)\delta(y-r)\right.\right.\\
&\left.+(t^{e}S(x)t^{a})^{il}S^{jm}(y)(t^{d}S(z))^{kn}\delta(z-r)\right]\delta(x-v)\delta(x-w)\\
&+\left[(t^{d}S(x)t^{a})^{il}(t^{e}S(y))^{jm}S^{kn}(z)\delta(x-r)+(S(x)t^{a})^{il}(t^{e}t^{d}S(y))^{jm}S^{kn}(z)\delta(y-r)\right.\\
&\left.+(S(x)t^{a})^{il}(t^{e}S(y))^{jm}(t^{d}S(z))^{kn}\delta(z-r)\right]\delta(y-v)\delta(x-w)\\
&+\left[(t^{d}S(x)t^{a})^{il}S^{jm}(y)(t^{e}S(z))^{kn}\delta(x-r)+(S(x)t^{a})^{il}(t^{d}S(y))^{jm}(t^{e}S(z))^{kn}\delta(y-r)\right.\\
&\left.\left.+(S(x)t^{a})^{il}S^{jm}(y)(t^{e}t^{d}S(z))^{kn}\delta(z-r)\right]\delta(z-v)\delta(x-w)\right)+\left(x\leftrightarrow y\right)+\left(x\leftrightarrow z\right)\,. \\
\end{split}\end{equation}
And the  $LRR$  term:
\begin{equation}\begin{split}
&f^{acb}J_{L}^{a}(w)S_{A}^{cd}(u)S_{A}^{be}(u^{\prime})J_{R}^{d}(r)J_{R}^{e}(v)B(x,y,z)=f^{ade}S_{A}^{dc}(u)S_{A}^{eb}(u^{\prime})\varepsilon^{ijk}\varepsilon^{lmn}\\
&\times \left(\left[(t^{a}S(x)t^{c}t^{b})^{il}S^{jm}(y)S^{kn}(z)\delta(x-r)+(t^{a}S(x)t^{b})^{il}(S(y)t^{c})^{jm}S^{kn}(z)\delta(y-r)\right.\right.\\
&\left.+(t^{a}S(x)t^{b})^{il}S^{jm}(y)(S(z)t^{c})^{kn}\delta(z-r)\right]\delta(x-w)\delta(x-v)\\
&+\left[(t^{a}S(x)t^{c})^{il}(S(y)t^{b})^{jm}S^{kn}(z)\delta(x-r)+(t^{a}S(x))^{il}(S(y)t^{c}t^{b})^{jm}S^{kn}(z)\delta(y-r)\right.\\
&\left.+(t^{a}S(x))^{il}(S(y)t^{b})^{jm}(S(z)t^{c})^{kn}\delta(z-r)\right]\delta(y-v)\delta(x-w)\\
&+\left[(t^{a}S(x)t^{c})^{il}S^{jm}(y)(S(z)t^{b})^{kn}\delta(x-r)+(t^{a}S(x))^{il}(S(y)t^{c})^{jm}(S(z)t^{b})^{kn}\delta(y-r)\right.\\
&\left.\left.+(t^{a}S(x))^{il}S^{jm}(y)(S(z)t^{c}t^{b})^{kn}\delta(z-r)\right]\delta(z-v)\delta(x-w)\right)+\left(x\leftrightarrow y\right)+\left(x\leftrightarrow z\right)\,.\\
\end{split}\end{equation}
Multiplying by  the kernel and integrating, we obtain for the connected part
 \begin{equation}\begin{split}
&\dot B_{JJSSJ}^{connected}\equiv \int_{r,v,u,u^{\prime},w}K_{JJSSJ}(w;r,v;u,u^{\prime})f^{acb}
\left[J_{L}^{d}(r)J_{L}^{e}(v)S_A^{dc}(u)S_A^{eb}(u^{\prime})J_{R}^{a}(w)\right. \\
&\left. \ \ \ \ \ \ \ \ \ \ \ \ \ \ \ \  \ \ \ \ \ \ \ \ \ \ \ \ \ \ -J_{L}^{a}(w)S_A^{cd}(u)S_A^{be}(u^{\prime})J_{R}^{d}(r)J_{R}^{e}(v)\right]B(x,y,z)\mid_{conn.}=\\
&=2 \int_{u,u^{\prime}}S_{A}^{dc}(u)S_{A}^{eb}(u^{\prime})\varepsilon^{ijk}\varepsilon^{lmn}K_{JJSSJ}(x;z,y;u,u^{\prime})\left(f^{acb}(S(x)t^{a})^{il}(t^{e}S(y))^{jm}(t^{d}S(z))^{kn}\right.\\
&\left.-f^{ade}(t^{a}S(x))^{il}(S(y)t^{b})^{jm}(S(z)t^{c})^{kn}\right)+\left(x\leftrightarrow y\right)+\left(x\leftrightarrow z\right)=\\
&=-\int_{u,u^{\prime}}\frac{i}{2}K_{JJSSJ}(x;z,y;u,u^{\prime})\left((S(x)S^{\dagger}(u^{\prime})S(y))\cdot(S(u^{\prime})S^{\dagger}(u)S(z))\cdot S(u)\right. \\
&\left. \ \ \ \ \ \ \ \ -(S(x)S^{\dagger}(u)S(z))\cdot(S(u)S^{\dagger}(u^{\prime})S(y))\cdot S(u^{\prime})\right.\\
&\left.+(S(y)S^{\dagger}(u^{\prime})S(x))\cdot(S(z)S^{\dagger}(u)S(u^{\prime}))\cdot S(u)-(S(z)S^{\dagger}(u)S(x))\cdot(S(y)S^{\dagger}(u^{\prime})S(u))\cdot S(u^{\prime})\right)\\
&+\left(x\leftrightarrow y\right)+\left(x\leftrightarrow z\right)\,. \\
  \end{split}\end{equation}
By  comparing with eq. $(4.28)$  of \cite{Grab}  we deduce the kernel $K_{JJSSJ}$ as quoted in eq.(\ref{KJJSJ}).

For the disconnected part we find
\begin{equation}\begin{split}
&\dot B_{JJSSJ}^{disconnected}\equiv\int_{r,v,u,u^{\prime},w}K_{JJSSJ}(w;r,v;u,u^{\prime})f^{acb}\left[J_{L}^{d}(r)J_{L}^{e}(v)S_A^{dc}(u)S_A^{eb}(u^{\prime})J_{R}^{a}(w)\right. \\
&\left. \ \ \ \ \ \ \ \ \ \ \ \ \ \ \  \ -J_{L}^{a}(w)S_A^{cd}(u)S_A^{be}(u^{\prime})J_{R}^{d}(r)J_{R}^{e}(v)\right]B(x,y,z)\mid_{disconn.}=\\
 &=-\int_{u,u^{\prime}}\Big\{ \frac{i}{4}K_{JJSSJ}(x,x,x,u,u^{\prime})\Big[(S(u^{\prime})S^{\dagger}(u)S(x)S^{\dagger}(u^{\prime})S(u))\cdot S(y)\cdot S(z) \\
&  -N_{c}^{2}B(u^{\prime},y,z)s(u,x)s(u^{\prime},u)\Big] 
-\frac{i}{2}K_{JJSSJ}(x,y,x,u,u^{\prime})\\
&\times  \Big[N_{c}s(u^{\prime},x)(S(u^{\prime})S^{\dagger}(u)S(y))\cdot S(u)\cdot S(z)+S(u^{\prime})\cdot(S(u)S^{\dagger}(u^{\prime})S(x)S^{\dagger}(u)S(y))\cdot S(z)\\
&+(S(y)S^{\dagger}(u)S(x)S^{\dagger}(u^{\prime})S(u))\cdot S(u^{\prime})\cdot S(z)+N_{c}s(u^{\prime},x)S(u)\cdot(S(y)S^{\dagger}(u)S(u^{\prime}))\cdot S(z)\Big]
\\
&+\frac{i}{4}K_{JJSSJ}(x,y,y,u,u^{\prime})  \\
&\times \Big[(S(x)S^{\dagger}(u^{\prime})S(u))\cdot(S(u^{\prime})S^{\dagger}(u)S(y))\cdot S(z)+N_{c}s(u^{\prime},u)(S(x)S^{\dagger}(u)S(y))\cdot S(u^{\prime})\cdot S(z)\\
&+(S(u)S^{\dagger}(u^{\prime})S(x))\cdot(S(y)S^{\dagger}(u)S(u^{\prime}))\cdot S(z)+N_{c}s(u^{\prime},u)(S(y)S^{\dagger}(u)S(x))\cdot S(u^{\prime})\cdot S(z)\Big] \\
&+(y\leftrightarrow z) \Big\} +\left(x\leftrightarrow y\right)+\left(x\leftrightarrow z\right)
 \end{split}\end{equation}


\beq
\dot B_{JJSSJ}=\dot B_{JJSSJ}^{connected}+\dot B_{JJSSJ}^{disconnected}\,.
\eeq

\subsection*{$\bf K_{JSJ}$}

The action of the $K_{JSJ}$ term is exactly of the leading order form (\ref{LOB}).
 For reasons that will become apparent in the next subsection, we write this term splitting the real and virtual contributions 
 \begin{equation}\begin{split}\label{BJSJ}
&\dot B_{JSJ}\equiv
\int_{v,w,u}K_{JSJ}(v,w;u)\left[J_{L}^{a}(v)J_{L}^{a}(w)+J_{R}^{a}(v)J_{R}^{a}(w)-2J_{L}^{a}(v)S_A^{ab}(u)J_{R}^{b}(w)\right]B(x,y,z)=\\
&=-{8\over 3} \int_{u}K_{JJ}(x,y;u)B(x,y,z)+  \int_{u}K_{JSJ}(x,y;u)\Big[-{1\over 3} B(x,y,z)+\frac{1}{2}\left(B(x,u,u)B(z,y,u)\right.\\ 
&\left.\ \ \ \ \ \ \ \ \ \ \ \ \ \ \ \ \ \ \ \ \ \ \ \ +B(y,u,u)B(z,x,u)-B(z,u,u)B(y,x,u)\right)\Big]+(x\leftrightarrow z)+(y\leftrightarrow z)\,,\\
 \end{split}\end{equation}
 where we have introduced separate notation for the kernel in the virtual term.
\beq
K_{JJ}(x,y;u)= K_{JSJ}(x,y;u)\,.\eeq

\subsection*{$\bf K_{JSSJ}$}
\begin{equation}\begin{split}
&f^{abc}f^{def}J_{L}^{a}(v)S_A^{be}(u)S_A^{cf}(u^{\prime})J_{R}^{d}(w)B(x,y,z)=4f^{abc}f^{def}\varepsilon^{ijk}\varepsilon^{lmn}\delta(x-w)\\
&\times \left[(t^{a}S(x)t^{d})^{il}S^{jm}(y)S^{kn}(z)\delta(x-v)+(S(x)t^{d})^{il}(t^{a}S(y))^{jm}S^{kn}(z)\delta(y-v)+\right.\\
&\left.+(S(x)t^{d})^{il}S^{jm}(y)(t^{a}S(z))^{kn}\delta(z-v)\right]\times tr\left[t^{b}S(u)t^{e}S^{\dagger}(u)\right]tr\left[t^{c}S(u^{\prime})t^{f}S^{\dagger}(u^{\prime})\right] \\
&+(x\leftrightarrow y)+(x\leftrightarrow z)=\frac{1}{4}\delta(x-w)\delta(x-v)\Big[s(u^{\prime},u)s(u,x)B(u^{\prime},y,z)\\
&-\varepsilon^{ijk}\varepsilon^{lmn}\left(S(u^{\prime})S^{\dagger}(u)S(x)S^{\dagger}(u^{\prime})S(u)\right)^{il}S^{jm}(y)S^{kn}(z)
+s(u,u^{\prime})s(u^{\prime},x)B(u,y,z)\\
&-\varepsilon^{ijk}\varepsilon^{lmn}\left(S(u)S^{\dagger}(u^{\prime})S(x)S^{\dagger}(u)S(u^{\prime})\right)^{il}S^{jm}(y)S^{kn}(z)\Big] 
+\frac{1}{4}\delta(x-w)\delta(y-v)\varepsilon^{ijk}\varepsilon^{lmn}\\
&\times\Big[\left(S(x)S^{\dagger}(u^{\prime})S(u)\right)^{il}(S(u^{\prime})S^{\dagger}(u)S(y))^{jm}S^{kn}(z)-s(u^{\prime},u)\left(S(x)S^{\dagger}(u)S(y)\right)^{im}\\
&\times S^{jl}(u^{\prime})S^{kn}(z)-s(u,u^{\prime})\left(S(x)S^{\dagger}(u^{\prime})S(y)\right)^{im}S^{jl}(u)S^{kn}(z)\\
&+\left(S(x)S^{\dagger}(u)S(u^{\prime})\right)^{il}\left(S(u)S^{\dagger}(u^{\prime})S(y)\right)^{jm}S^{kn}(z)\Big]+\frac{1}{4}\delta(x-w)\delta(z-v)\varepsilon^{ijk}\varepsilon^{lmn}\\
&\times \Big[\left(S(x)S^{\dagger}(u^{\prime})S(u)\right)^{il}(S(u^{\prime})S^{\dagger}(u)S(z))^{kn}S^{jm}(y)-s(u^{\prime},u)\left(S(x)S^{\dagger}(u)S(z)\right)^{in}\\
&\times S^{kl}(u^{\prime})S^{jm}(y)-s(u,u^{\prime})\left(S(x)S^{\dagger}(u^{\prime})S(z)\right)^{im}S^{kn}(u)S^{jm}(y)\\
&+\left(S(x)S^{\dagger}(u)S(u^{\prime})\right)^{il}\left(S(u)S^{\dagger}(u^{\prime})S(z)\right)^{kn}S^{jm}(y)\Big]+(x\leftrightarrow y)+(x\leftrightarrow z)\,. \\
\end{split}\end{equation}
After insertion and integration we find:
\begin{equation}\begin{split}
&
\dot B_{JSSJ}\equiv
 \int_{v,w,u,u^{\prime}}K_{JSSJ}(v,w,u,u^{\prime})\left[f^{abc}f^{def}J_{L}^{a}(v)S_A^{be}(u)S_A^{cf}(u^{\prime})J_{R}^{d}(w)\right. \\
&\left. -N_{c}J_{L}^{a}(v)S_A^{ab}(u)J_{R}^{b}(w)\right]B(x,y,z)=\\
&=2 \int_{u,u^{\prime}} K_{JSSJ}(y,x,u,u^{\prime})\left[(S(x)S^{\dagger}(u^{\prime})S(u))\cdot(S(u^{\prime})S^{\dagger}(u)S(y))\cdot S(z)  \right.\\
&\left. + s(u^{\prime},u)(S(x)S^{\dagger}(u)S(y))\cdot S(u^{\prime})\cdot S(z)- (u^\prime\rightarrow u)\right] +(x\leftrightarrow z)+(y\leftrightarrow z)\,.\\
\end{split}\end{equation}

\subsection*{$\bf K_{q\bar{q}}$}
Finally, the quark contribution is: 
\begin{equation}\begin{split}
&2J_{L}^{a}(v)tr\left(S^{\dagger}(u)t^{a}S(u^{\prime})t^{b}\right)J_{R}^{b}(w)B(x,y,z)=
2\varepsilon^{ijk}\varepsilon^{lmn}\delta(x-w)\\
&\times \left[(t^{a}S(x)t^{b})^{il}S^{jm}(y)S^{kn}(z)\delta(x-v)+(S(x)t^{b})^{il}(t^{a}S(y))^{jm}S^{kn}(z)\delta(y-v)\right.\\
&\left.+(S(x)t^{b})^{il}S^{jm}(y)(t^{a}S(z))^{kn}\delta(z-v)\right]\times tr\left[t^{a}S(u^{\prime})t^{b}S^{\dagger}(u)\right]+\left(x\leftrightarrow y\right)+\left(x\leftrightarrow z\right)=\\
&=\frac{1}{2}\delta(x-w)\left[\left(N_{c}s(u,x)B(u^{\prime},y,z)-\frac{1}{N_{c}}(S(u^{\prime})S^{\dagger}(u)S(x))\cdot S(y)\cdot S(z)\right. \right. \\
&-\frac{1}{N_{c}}(S(x)S^{\dagger}(u)S(u^{\prime}))\cdot S(y)\cdot S(z)\left.+\frac{1}{N_{c}}B(x,y,z)s(u,u^{\prime})\right)\delta(x-v)\\
&+\left(-(S(x)S^{\dagger}(u)S(y))\cdot S(u^{\prime})\cdot S(z)-\frac{1}{N_{c}}(S(u^\prime)S^{\dagger}(u)S(y))\cdot S(x)\cdot S(z)\right.\\
&\left.-\frac{1}{N_{c}}(S(x)S^{\dagger}(u)S(u^{\prime}))\cdot S(y)\cdot S(z)+\frac{1}{N_{c}}B(x,y,z)s(u,u^{\prime})\right)\delta(y-v)\\
&+\left(-(S(x)S^{\dagger}(u)S(z))\cdot S(y)\cdot S(u^{\prime})-\frac{1}{N_{c}}(S(u^\prime)S^{\dagger}(u)S(z))\cdot S(x)\cdot S(y)\right.\\
&\left. -\frac{1}{N_{c}}(S(x)S^{\dagger}(u)S(u^{\prime}))\cdot S(y)\cdot S(z) \left.+\frac{1}{N_{c}}B(x,y,z)s(u,u^{\prime})\right)\delta(z-v)\right]+(x\leftrightarrow y)+(x\leftrightarrow z)\,.
\end{split}\end{equation}
After the integration we find:
\begin{equation}\begin{split}
&
\dot B_{qq}\equiv
\int_{_{v,u,u^{\prime},w}}K_{q\bar{q}}(v,w;u,u^{\prime})\left[2J_{L}^{a}(v)tr\left(S^{\dagger}(u)t^{a}S(u^{\prime})t^{b}\right)J_{R}^{b}(w)\right.\\
&\left. \ \ \ \ \ \ \ \ \ -J_{L}^{a}(v)S_A^{ab}(u)J_{R}^{b}(w)\right]B(x,y,z)=\\
&=\frac{1}{2}\int_{_{u,u^{\prime}}}K_{q\bar{q}}(y,x;u,u^{\prime})\Big\{
-(S(x)S^{\dagger}(u)S(y))\cdot S(u^\prime)\cdot S(z) 
-\frac{1}{3}(S(u^{\prime})S^{\dagger}(u)S(y))\cdot S(x)\cdot S(z) \\
&-\frac{1}{3}(S(x)S^{\dagger}(u)S(u^{\prime}))\cdot S(y)\cdot S(z)
+\frac{2}{3}B(x,y,z)s(u,u^{\prime})
-(S(y)S^{\dagger}(u)S(x))\cdot S(u^\prime)\cdot S(z) \\
&-\frac{1}{3}(S(u^{\prime})S^{\dagger}(u)S(x))\cdot S(y)\cdot S(z) 
-\frac{1}{3}(S(y)S^{\dagger}(u)S(u^{\prime}))\cdot S(x)\cdot S(z)
+3s(u,x)B(u,y,z) \\
& +3s(u,y)B(u,x,z)-3s(u,z)B(u,y,x)-\frac{1}{3}B(x,y,z)\Big\} +(x\leftrightarrow z)+(y\leftrightarrow z)\,.\\
\end{split}\end{equation}

\subsection{NLO evolution of conformal operator in ${\cal N}=4$ }
To complete the discussion of the evolution of the baryon operator, we present the evolution of the conformal baryon in ${\cal N}=4$ super Yang-Mills theory.
As mentioned in the introduction, the status of conformal invariance in JIMWLK evolution in ${\cal N}=4$  supersymmetric Yang-Mills theory has been understood in \cite{KLMconf}. It was shown there that the JIMWLK Hamiltonian at NLO posesses exact conformal symmetry, although the action of the conformal transformation on Wilson line operators is modified at NLO from the naive form. We have also constructed the modified baryon operator, which transforms under the naive conformal transformation.
The conformal extension ${\cal B}$ of the operator $B$ according to \cite{KLMconf} is given by:
\begin{eqnarray}
&&{\cal B}(u,v,w)=B(u,v,w)+\frac{3}{4}\int_z \left\{ M_{u,v,z}\ln\frac{(u-v)^2a^2}{(u-z)^2(v-z)^2} \Big[\frac{1}{ 6}(B(u,z,z)B(w,v,z)+\right. \nonumber \\ &&+B(v,z,z)B(w,u,z)-B(w,z,z)B(v,u,z)) 
 -\left. B(u,v,w)\Big]+(u\leftrightarrow w)+(v\leftrightarrow w)\right\}\,. \nonumber \\
\end{eqnarray}
The procedure of contructing the conformal extension as discussed in \cite{KLMconf} is very general and can be applied to any operator. In fact it is simply equivalent to a unitary transformation, or change of basis. When applied to the JIMWLK Hamlitonian it yields a unitarily equivalent ``conformal Hamiltonian'' $H^{NLO\ JIMLWK}_{conf}$, which is invariant under the naive conformal transformation. The evolution of ``conformal operators'' is then derived by the application of $H^{NLO\ JIMLWK}_{conf}$. The form of $H^{NLO\ JIMLWK}_{conf}$ is the same asa that of $H^{NLO\ JIMLWK}$, except that the expressions for various kernels are modified.
In particular, in the ${\cal N}=4$ super Yang Mills theory, as derived in \cite{KLMconf} the ``conformal kernels'' are

\begin{eqnarray}\label{k1}
&& {\cal K}_{JJSSJ}(w;x,y;z,z^\prime)
 =\frac{i}{ 4}\Big[ M_{x,y,z}M_{y,z,z^\prime}\ln\frac{W^4X^{\prime\,2}Y^{\prime\,2}}{W^{\prime\,4}X^2Y^2}
 +M_{x,w,z}M_{y,w,z^\prime} \ln\frac{(x-w)^2 W^2Y^{\prime\,2}}{(y-w)^2W^{\prime\,2}X^2} \nonumber \\
&& - M_{y,w,z^\prime} M_{x,z^\prime,z}\ln\frac{ W^4 X^{\prime\,2}Y^{\prime\,2}}{(y-w)^2 (z-z^\prime)^2W^{\prime\,2}X^2}
 -M_{x,w,z}M_{y,z,z^\prime}  \ln\frac{ (x-w)^2 (z-z^\prime)^2 W^2 Y^{\prime\,2}}{W^{\prime\,4}X^2Y^2}
    \Big]; \nonumber\\ \nonumber \\
 && {\cal K}_{JJSJ}(w;x,y;z)=  \int_{z^\prime}\, \left[{\cal K}_{JJSSJ}(y;w,x;z,z^\prime)\,-\,{\cal K}_{JJSSJ}(x;w,y;z,z^\prime)\right];
\nonumber \\ \nonumber \\
  && {\cal K}_{JSSJ}(x,y;z,z^\prime)= \frac{\alpha_s^2}{16\,\pi^4}
\Bigg[\frac{(x-y)^2}{ X^2Y'^2(z-z^\prime)^2}\Big(1+\frac{(x-y)^2(z-z')^2}{X^2Y'^2-X'^2Y^2}\Big)-\nonumber \\
&&~~~~~~~~~~~~~~~~~~~~~~~~~~~~~~-\frac{(x-y)^2}{ X'^2Y^2(z-z^\prime)^2}\Big(1+\frac{(x-y)^2(z-z')^2}{X'^2Y^2-X^2Y'^2}\Big)\Bigg]\ \ln\frac{X^2{Y'}^2}{ {X'}^2Y^2}\nonumber \\
  &&+ \frac{\alpha_s^2}{16\pi^4}\frac{(x-y)^2}{(z-z^\prime)^2}
  \left[ \frac{1}{Y^2X^{\prime\,2}} \ln \frac{(x-y)^2(z-z^\prime)^2}{X^2Y^{\prime\,2}} 
  + \frac{1}{X^2Y^{\prime\,2}} \ln \frac{(x-y)^2(z-z^\prime)^2}{Y^2X^{\prime\,2}} \right];
  \nonumber \\ \nonumber \\
  && {\cal K}_{JSJ}(x,y;z)= \frac{\alpha_s^2N_c}{48 \pi}\frac{(x-y)^2}{X^2 Y^2}\nonumber \\ 
  &&- \frac{\alpha_s^2\,N_c}{16\pi^4}\int_{z^\prime} \frac{(x-y)^2}{(z-z^\prime)^2}
  \left[ \frac{1}{Y^2X^{\prime\,2}} \ln \frac{(x-y)^2(z-z^\prime)^2}{X^2Y^{\prime\,2}} 
  + \frac{1}{X^2Y^{\prime\,2}} \ln \frac{(x-y)^2(z-z^\prime)^2}{Y^2X^{\prime\,2}} \right].
  \nonumber  \\ \nonumber \\
&& {\cal K}_{JJ}(x,y,z)\ =\ \frac{\alpha_s^2N_c}{48 \pi}\frac{(x-y)^2}{X^2 Y^2} .
\end{eqnarray}
And, obviously, ${\cal K}_{qq}=0$.

In ${\cal N}=4$, the evolution of $\cal B$ is therefore given by the same equation (\ref{NLOB}) but with  the kernels $K$ replaced by the conformal kernels $\cal K$.
Linearized NLO evolution equation for ${\cal B}$ should coincide with the result of ref. \cite{NLOBKP}, but this comparison is beyond the scope of the present paper.

\appendix

\section{Action of the NLO JIMWLK Hamiltonian on a dipole}

Ref. \cite{BC}  has computed the evolution of a quark-antiquark dipole $s$ at NLO.  We quote this evolution equation (eq. (5) of ref. \cite{BC}) for the sake of self-consistency of presentation
\begin{equation}\begin{split}&\frac{d}{dY}s(u,v)=\\
&=\frac{\alpha_{s}N_{c}}{2\pi^{2}}\int d^{2}z\frac{(u-v)^{2}}{U^{2}V^{2}}\left\{ 1+\frac{\alpha_{s}}{4\pi}\left[b\ln(u-v)^{2}\mu^{2}-b\frac{U^{2}-V^{2}}{(u-v)^{2}}\ln\frac{U^{2}}{V^{2}}+
\left(\frac{67}{9}-\frac{\pi^{3}}{3}\right)N_{c}\right.\right.\\
&\left.\left.-\frac{10}{9}n_{f}-2N_{c}\ln\frac{U^{2}}{(u-v)^{2}}\ln\frac{V^{2}}{(u-v)^{2}}\right]\right\} \left[s(u,z)s(z,v)-s(u,v)\right]\\
&+\frac{\alpha_{s}^{2}}{16N_{c}\pi^{4}}\int d^{2}zd^{2}z^{\prime}\left[\left(-\frac{4}{(z-z^{\prime})^{4}}+\left\{ 2\frac{U^{2}(V^{\prime})^{2}+(U^{\prime})^{2}V^{2}-4(u-v)^{2}(z-z^{\prime})^{2}}{(z-z^{\prime})^{2}\left[U^{2}(V^{\prime})^{2}-(U^{\prime})^{2}V^{2}\right]}\right.\right.\right.\\
&\left.\left.+\frac{(u-v)^{4}}{U^{2}(V^{\prime})^{2}-(U^{\prime})^{2}V^{2}}\left[\frac{1}{U^{2}(V^{\prime})^{2}}+\frac{1}{V^{2}(U^{\prime})^{2}}\right]+\frac{(u-v)^{2}}{(z-z^{\prime})^{2}}\left[\frac{1}{U^{2}(V^{\prime})^{2}}-\frac{1}{(U^{\prime})^{2}V^{2}}\right]\right\} \right.  \\ 
&\left. \times  \ln\frac{U^{2}(V^{\prime})^{2}}{(U^{\prime})^{2}V^{2}}\right)
\left[N_{c}^{3}s(u,z^{\prime})s(z^{\prime},z)s(z,v)-Tr\left(S^{\dagger}(u)S(z)S^{\dagger}(z^{\prime})S(v)S^{\dagger}(z)S(z^{\prime})\right)-(z^{\prime}\rightarrow z)\right]\\
&-\left\{ \frac{(u-v)^{2}}{(z-z^{\prime})^{2}}\left[\frac{1}{U^{2}(V^{\prime})^{2}}+\frac{1}{V^{2}(U^{\prime})^{2}}\right]-\frac{(u-v)^{4}}{V^{2}(U^{\prime})^{2}(V^{\prime})^{2}U^{2}}\right\} \ln\left(\frac{U^{2}(V^{\prime})^{2}}{(U^{\prime})^{2}V^{2}}\right)  \\
&\times N_{c}^{3}s(u,z^{\prime})s(z^{\prime},z)s(z,v)
+4n_{f}\left\{ \frac{4}{(z-z^{\prime})^{4}}-2\frac{U^{2}(V^{\prime})^{2}+(U^{\prime})^{2}V^{2}-(u-v)^{2}(z-z^{\prime})^{2}}{(z-z^{\prime})^{4}((V^{\prime})^{2}U^{2}-V^{2}(U^{\prime})^{2})}\right. \\ 
&\left. \times \ln\left(\frac{U^{2}(V^{\prime})^{2}}{(U^{\prime})^{2}V^{2}}\right)\right\} Tr\left\{ S^{\dagger}(u)t^{a}S(v)t^{b}\right\} 
\left.\left[Tr\left\{ t^{a}S(z)t^{b}S^{\dagger}(z^{\prime})\right\} -(z^{\prime}\rightarrow z)\right]\right]\,.
\label{dipole}
\end{split}\end{equation}
Here we used the notations of \cite{BC}:
\begin{equation}\begin{split}
U=u-z;\qquad\qquad U^{\prime}=u-z^{\prime};\qquad\qquad
V=v-z;\qquad\qquad V^{\prime}=v-z^{\prime}.
\end{split}\end{equation} 

In the JIMWLK formalism, this
evolution is generated by applying $H^{NLO\,JIMWLK}$ to $s(u,v)$ according to eq.(\ref{1}). The action is defined through the action of the rotation generators $J_L$ 
and $J_R$ (\ref{LR}) and is a purely algebraic operation. It is easy to see that all five kernels contribute to the evolution of the dipole and each contribution can be identified in eq.(\ref{dipole}).

We now list the action  of every term in the Hamiltonian (labeled by its kernel) on the dipole.                  

\subsection*{$\ensuremath{\mathbf{K_{JSJ}}}$}
Action of the rotation operators on  $s(u,v)$:
\begin{equation}\begin{split}&J_{L}^{a}(x)J_{L}^{a}(y)s(u,v)=\frac{1}{N_{c}}tr\left[S^{\dagger}(u)t^{a}t^{a}S(v)\right]\left[\delta(v-x)-\delta(u-x)\right]\left[\delta(v-y)-\delta(u-y)\right]\\
&=\frac{N^{2}_{c}-1}{2N^{2}_{c}}tr\left[S^{\dagger}(u)S(v)\right]\left[\delta(v-x)-\delta(u-x)\right]\left[\delta(v-y)-\delta(u-y)\right]\\
&=\frac{N^{2}_{c}-1}{2N_{c}}\left[\delta(v-x)-\delta(u-x)\right]\left[\delta(v-y)-\delta(u-y)\right]s(u,v)\\
&J_{R}^{a}(x)J_{R}^{a}(y)s(u,v)=\frac{1}{N_{c}}tr\left[S^{\dagger}(u)S(v)t^{a}t^{a}\right]\left[\delta(v-x)-\delta(u-x)\right]\left[\delta(v-y)-\delta(u-y)\right]\\
&=\frac{N^{2}_{c}-1}{2N_{c}}\left[\delta(v-x)-\delta(u-x)\right]\left[\delta(v-y)-\delta(u-y)\right]s(u,v)\\
&J_{L}^{a}(x)J_{R}^{d}(y)s(u,v)=\frac{1}{N_{c}}tr\left[S^{\dagger}(u)t^{a}S(v)t^{d}\right]\left[\delta(v-y)-\delta(u-y)\right]\left[\delta(v-x)-\delta(u-x)\right]\,.
\label{jj}
\end{split}\end{equation}
\begin{eqnarray}
&&2J_{L}^{a}(x)S_A^{ab}(z)J_{R}^{b}(y)s(u,v)=\frac{4}{N_{c}}tr\left[t^{a}S(z)t^{b}S^{\dagger}(z)\right]tr\left[S^{\dagger}(u)t^{a}S(v)t^{b}\right]\left[\delta(v-x)-\delta(u-x)\right]
\nonumber \\ && \times \left[\delta(v-y)-\delta(u-y)\right]\nonumber \\
&&=\left[\delta(v-x)-\delta(u-x)\right]\left[\delta(v-y)-\delta(u-y)\right]\left[N_{c}s(u,z)s(z,v)-\frac{1}{N_{c}}s(u,v)\right]\,,
\end{eqnarray}
where we have used the identity
 \beq S_{A}^{ab}(z)=2tr\left[t^{a}S(z)t^{b}S^{\dagger}(z)\right]\label{adjoint}.\eeq
 Finally, the action of the "$K_{JSJ}$" term in the Hamiltonian  reads 
\begin{equation}\begin{split}&\int_{x,y,z}K_{JSJ}(x,y;z)\left[J_{L}^{a}(x)J_{L}^{a}(y)+J_{R}^{a}(x)J_{R}^{a}(y)-2J_{L}^{a}(x)S_A^{ab}(z)J_{R}^{b}(y)\right]s(u,v)\\
&=N_{c}\int_{z}\left[K_{JSJ}(v,v;z)+K_{JSJ}(u,u;z)-K_{JSJ}(u,v;z)-K_{JSJ}(v,u;z)\right]\\
&\times \left[s(u,v)-s(u,z)s(z,v)\right]=2N_{c}\int_{z}K_{JSJ}(u,v;z)\left[s(u,z)s(z,v)-s(u,v)\right]\,,
\end{split}\end{equation}
To get the last equality we have used the fact that the kernel is symmetric: $K_{JSJ}(u,v;z)=K_{JSJ}(v,u;z)$ and $K_{JSJ}(u,u;z)=0$.
Obviously, we just recover the leading order dipole evolution if instead of $K_{JSJ}$ we take $-M/2$.

\subsection*{$\mathbf{K_{JSSJ}}$}
 Below we use  the completeness relation 
 \beq\label{completness}
 t_{\alpha\beta}^{a}t_{\gamma\delta}^{a}=\frac{1}{2}\left[\delta_{\alpha\delta}\delta_{\beta\gamma}-\frac{1}{N_{c}}\delta_{\alpha\beta}\delta_{\gamma\delta}\right]\,,
 \eeq
 and the following identity :
 \begin{equation}
f^{abc}\, t_{jk}^{a}\, t_{mn}^{b}\, t_{qu}^{c}
=\frac{i}{4}\left(\delta_{ju}\delta_{qn}\delta_{mk}-\delta_{jn}\delta_{mu}\delta_{kq}\right)\label{ttt}\ ,
 \end{equation}
 obtained using $f^{abc}t_{qu}^{c}=-i\left(t_{qd}^{a}t_{du}^{b}-t_{qd}^{b}t_{du}^{a}\right)$.
 
Using eq. (\ref{jj}) and the identity eq. (\ref{ttt}) we obtain
\begin{equation}\begin{split}
&\int_{x\, y\, z\, z^{\prime}}K_{JSSJ}(x,y;z,z^{\prime})\left[f^{abc}f^{def}J_{L}^{a}(x)S_A^{be}(z)S_A^{cf}(z^{\prime})J_{R}^{d}(y)\right]s(u,v)=\\
&=\frac{1}{4}\int_{x\, y\, z\, z^{\prime}}\,\left(K_{JSSJ}(v,v;z,z^{\prime})-K_{JSSJ}(u,v;z,z^{\prime})-K_{JSSJ}(v,u;z,z^{\prime})+K_{JSSJ}(u,u;z,z^{\prime})\right)\\
&\times \left[N_{c}^{2}s(u,z^{\prime})s(z^{\prime},z)s(z,v)+N_{c}^{2}s(u,z)s(z,z^{\prime})s(z^{\prime},v)-\frac{1}{N_{c}}tr\left[S^{\dagger}(u)S(z^{\prime})S^{\dagger}(z)S(v)S^{\dagger}(z^{\prime})S(z)\right]\right.\\
&\left.-\frac{1}{N_{c}}tr\left[S^{\dagger}(u)S(z)S^{\dagger}(z^{\prime})S(v)S^{\dagger}(z)S(z^{\prime})\right]\right]\\
&=\frac{1}{2}\int_{z\, z^{\prime}}\,\left(K_{JSSJ}(u,u;z,z^{\prime})+K_{JSSJ}(v,v;z,z^{\prime})-K_{JSSJ}(v,u;z,z^{\prime})-K_{JSSJ}(u,v;z,z^{\prime})\right)\\
&\times \left[N_{c}^{2}s(u,z^{\prime})s(z^{\prime},z)s(z,v)-\frac{1}{N_{c}}tr\left[S^{\dagger}(u)S(z)S^{\dagger}(z^{\prime})S(v)S^{\dagger}(z)S(z^{\prime})\right]\right]\\
&=-\int_{z\, z^{\prime}}\, K_{JSSJ}(u,v;z,z^{\prime})\left[N_{c}^{2}s(u,z^{\prime})s(z^{\prime},z)s(z,v)-\frac{1}{N_{c}}tr\left[S^{\dagger}(u)S(z)S^{\dagger}(z^{\prime})S(v)S^{\dagger}(z)S(z^{\prime})\right]\right]\\
\end{split}\end{equation}
In the last equality we have again used the symmetry of the kernel: $K_{JSSJ}(u,v,z,z^{\prime})=K_{JSSJ}(v,u,z,z^{\prime})$ and  $K_{JSSJ}(u,u,z,z^{\prime})=0$.

\subsection*{$\mathbf{K_{q\bar{q}}}$}

The contribution from the quarks can be written as:
\begin{equation}\begin{split}
&\int_{x,y,z,z^{\prime}}K_{q\bar{q}}(x,y;z,z^{\prime})\left[2J_{L}^{a}(x)tr\left[S^{\dagger}(z)t^{a}S(z^{\prime})t^{b}\right]J_{R}^{b}(y)\right]s(u,v)\\
&=\frac{1}{N_{c}}\int_{z\, z^{\prime}}\,\left(K_{q\bar{q}}(u,u;z,z^{\prime})+K_{q\bar{q}}(v,v;z,z^{\prime})-K_{q\bar{q}}(v,u;z,z^{\prime})-K_{q\bar{q}}(u,v;z,z^{\prime})\right)\\
&\times 2tr\left[S^{\dagger}(u)t^{a}S(v)t^{b}\right]tr\left[S^{\dagger}(z)t^{a}S(z^{\prime})t^{b}\right]\\
&=\frac{1}{2N_{c}}\int_{z\, z^{\prime}}\,\left(K_{q\bar{q}}(u,u;z,z^{\prime})+K_{q\bar{q}}(v,v;z,z^{\prime})-K_{q\bar{q}}(v,u;z,z^{\prime})-K_{q\bar{q}}(u,v;z,z^{\prime})\right)\\
&\times \Big(N_{c}s(u,z^{\prime})s(z,v)-\frac{1}{N_{c}^{2}}tr\left[S^{\dagger}(u)S(v)S^{\dagger}(z)S(z^{\prime})\right]-\frac{1}{N_{c}^{2}}tr\Big[S^{\dagger}(u)S(v)S^{\dagger}(z^{\prime})S(z)\Big]\\ &+\frac{1}{N_{c}}s(u,v)s(z,z^{\prime})\Big)= -\int_{z\, z^{\prime}}\, K_{q\bar{q}}(u,v;z,z^{\prime})\times\\
&\times\Big(N_{c}s(u,z^{\prime})s(z,v)-\frac{1}{N_{c}^{2}}tr\left[S^{\dagger}(u)S(v)S^{\dagger}(z)S(z^{\prime})\Big]-\frac{1}{N_{c}^{2}}tr\Big[S^{\dagger}(u)S(v)S^{\dagger}(z^{\prime})S(z)\right]\\ & +\frac{1}{N_{c}}s(u,v)s(z,z^{\prime})\Big)\,.
\end{split}\end{equation}
$K_{q\bar{q}}(u,v;z,z^{\prime})=K_{q\bar{q}}(v,u;z,z^{\prime})$ and  $K_{q\bar{q}}(u,u;z,z^{\prime})=0$ were used.

\subsection*{$\mathbf{K_{JJSJ}}$}
We first calculate the action of $J_LJ_LJ_R$ on $s(u,v)$ is:
  \begin{eqnarray}\label{LLR}
&&J_{L}^{d}(x)J_{L}^{e}(y)J_{R}^{a}(w)s(u,v)=\nonumber \\
&&\ \ \ \ =\frac{1}{N_{c}}\left[\delta(v-w)-\delta(u-w)\right]\left[\delta(v-y)-\delta(u-y)\right]\Big[tr\Big[S^{\dagger}(u)t^{e}t^{d}S(v)t^{a}\Big]\delta(v-x)\nonumber \\ 
&&\ \ \ \ -tr\Big[S^{\dagger}(u)t^{d}t^{e}S(v)t^{a}\Big]\delta(u-x)\Big]\,.
\end{eqnarray}
Contraction with the representation (\ref{adjoint}) of the adjoint Wilson line gives:
 \begin{equation}\begin{split}
&2f^{bde}tr\left(t^{b}S(z)t^{a}S^{\dagger}(z)\right)tr\left(S^{\dagger}(u)t^{e}t^{d}S(v)t^{a}\right)=2f^{bde}t_{ij}^{b}S_{jk}(z)t_{kl}^{a}S_{li}^{\dagger}(z)S_{rs}^{\dagger}(u)t_{st}^{e}t_{tm}^{d}S_{mp}(v)t_{pr}^{a}\\
&=2f^{bde}t_{ij}^{b}t_{tm}^{d}t_{st}^{e}t_{kl}^{a}t_{pr}^{a}S_{jk}(z)S_{li}^{\dagger}(z)S_{rs}^{\dagger}(u)S_{mp}(v)=\frac{i}{4}\left(\delta_{ij}\delta_{ms}-N_{c}\delta_{js}\delta_{im}\right)\left(\delta_{kr}\delta_{pl}-\frac{1}{N_{c}}\delta_{kl}\delta_{pr}\right) \\ 
& \times S_{jk}(z)S_{li}^{\dagger}(z)S_{rs}^{\dagger}(u)S_{mp}(v)\\
&=\frac{i}{4}\left(N_{c}s(u,v)-s(u,v)-N_{c}^{3}s(u,z)s(z,v)+s(u,v)\right)=\frac{i}{4}N_{c}\left(s(u,v)-N_{c}^{2}s(u,z)s(z,v)\right)\,.
\end{split}\end{equation}
Thus we find:
  \begin{equation}\begin{split}
&f^{bde}J_{L}^{d}(x)J_{L}^{e}(y)S_A^{ba}(z)J_{R}^{a}(w)s(u,v)=\frac{i}{4}\left[\delta(v-w)-\delta(u-w)\right] \\
&\times \left[\delta(v-y)-\delta(u-y)\right]\left[\delta(v-x)+\delta(u-x)\right]\left[s(u,v)-N_{c}^{2}s(u,z)s(z,v)\right]
\end{split}\end{equation}
Similarly for the  $J_RJ_RJ_L$ term
  \begin{equation}\begin{split}
&-f^{bde}J_{L}^{a}(w)S_A^{ab}(z)J_{R}^{d}(x)J_{R}^{e}(y)s(u,v)=\frac{i}{4}\left[\delta(u-w)-\delta(v-w)\right]\\
&\times \left[\delta(u-y)-\delta(v-y)\right]\left[\delta(u-x)+\delta(v-x)\right]\left[s(u,v)-N_{c}^{2}s(z,v)s(u,z)\right]\,.\\
\end{split}\end{equation}
Adding both contributions together, multiplying by the kernel $K_{JJSJ}$  and performing the integrations:
  \begin{eqnarray}
&&\int_{w,x,y,z}K_{JJSJ}(w;x,y;z)f^{bde}\left[J_{L}^{d}(x)J_{L}^{e}(y)S_A^{ba}(z)J_{R}^{a}(w)-J_{L}^{a}(w)S_A^{ab}(z)J_{R}^{d}(x)J_{R}^{e}(y)\right]s(u,v)\nonumber \\
&&=\frac{i}{4}\int_{z}\Big[K_{JJSJ}(v;v,v;z)+K_{JJSJ}(v;u,v;z)-K_{JJSJ}(v;v,u;z)-K_{JJSJ}(v;u,u;z)\nonumber\\
&&-K_{JJSJ}(u;v,v;z)-K_{JJSJ}(u;u,v;z)+K_{JJSJ}(u;v,u;z)+K_{JJSJ}(u;u,u;z)\Big ]\nonumber\\
&&\times \left[s(u,v)-N_{c}^{2}s(u,z)s(z,v)\right]+\frac{i}{4}\int_{z} [K_{JJSJ}(u;u,u;z)+K_{JJSJ}(u;v,u;z)\nonumber \\ 
&&-K_{JJSJ}(u;u,v;z)-K_{JJSJ}(u;v,v;z)-K_{JJSJ}(v;u,u;z)-K_{JJSJ}(v;v,u;z)\nonumber\\
&&+K_{JJSJ}(v;u,v;z)+K_{JJSJ}(v;v,v;z)]\left[s(u,v)-N_{c}^{2}s(u,z)s(z,v)\right]\nonumber\\
&&=\frac{i}{2}\int_{z}\left.[
K_{JJSJ}(v;u,v;z)-K_{JJSJ}(v;v,u;z)
-K_{JJSJ}(u;u,v;z)+K_{JJSJ}(u;v,u;z)
\right]\nonumber \\
&&\times \left[s(u,v)-N_{c}^{2}s(u,z)s(z,v)\right]\label{rJJSJ}\,.
 \end{eqnarray}

We now discuss the action of the virtual terms.
\begin{eqnarray}\label{LLL}
&&f^{bde}J_{L}^{d}(x)J_{L}^{e}(y)J_{L}^{b}(w)s(u,v)=\nonumber \\
&&=-\frac{f^{bde}}{N_{c}}\left[tr\left[S^{\dagger}(u)t^{d}t^{b}t^{e}S(v)\right]\delta(v-y)\delta(u-x)-tr\left[S^{\dagger}(u)t^{b}t^{e}t^{d}S(v)\right]\delta(v-y)\delta(v-x)\right.\nonumber \\
&&\left.+tr\left[S^{\dagger}(u)t^{e}t^{b}t^{d}S(v)\right]\delta(u-y)\delta(v-x)-tr\left[S^{\dagger}(u)t^{d}t^{e}t^{b}S(v)\right]\delta(u-y)\delta(u-x)\right]\nonumber \\
&&\times \left[\delta(v-w)-\delta(u-w)\right]=\nonumber \\
&&-\frac{N_{c}^{2}-1}{4}i\left[-\delta(v-y)\delta(u-x)+\delta(v-y)\delta(v-x)+\delta(u-y)\delta(v-x)-\delta(u-y)\delta(u-x)\right]\nonumber \\
&&\times \left[\delta(v-w)-\delta(u-w)\right]s(u,v)
 \end{eqnarray}
 Similarly  the $J_RJ_RJ_R$ term
 \begin{eqnarray}\label{RRR}
&&f^{bde}J_{R}^{d}(x)J_{R}^{e}(y)J_{R}^{b}(w)s(u,v)=\nonumber \\
&&=\frac{f^{bde}}{N_{c}}\left[tr\left[S(v)t^{d}t^{b}t^{e}S^{\dagger}(u)\right]\delta(u-y)\delta(v-x)-tr\left[S(v)t^{b}t^{e}t^{d}S^{\dagger}(u)\right] \delta(u-y)\delta(u-x)\right.\nonumber \\
&&\left.+tr\left[S(v)t^{e}t^{b}t^{d}S^{\dagger}(u)\right]\delta(v-y)\delta(u-x)-tr\left[(S(v)t^{d}t^{e}t^{b}S^{\dagger}(u)\right]\delta(v-y)\delta(v-x)\right]\nonumber \\
&&\times \left[\delta(u-w)-\delta(v-w)\right]s(u,v)=\nonumber \\
&&-\frac{N_{c}^{2}-1}{4}i\left[-\delta(u-y)\delta(v-x)+\delta(u-y)\delta(u-x)+\delta(v-y)\delta(u-x)-\delta(v-y)\delta(v-x)\right]\nonumber \\
&&\left[\delta(u-w)-\delta(v-w)\right]s(u,v)\,.
\end{eqnarray}
After the integrations the virtual term has the form:
\begin{equation}\begin{split}
&\frac{1}{3}\int_{w,x,y,z}K_{JJSJ}(w;x,y;z)f^{bde}\left[J_{L}^{d}(x)J_{L}^{e}(y)J_{L}^{b}(w)-J_{R}^{d}(x)J_{R}^{e}(y)J_{R}^{b}(w)\right]s(u,v)\\
&=i\frac{N_{c}^{2}-1}{3}\int_{z}[K_{JJSJ}(u,v,u,z)+K_{JJSJ}(v,u,v,z) ]s(u,v)\,. \\
 \end{split}\end{equation}
In the above we used the antisymmetry of the kernel $K_{JJSJ}(w,x,y;z)=-K_{JJSJ}(w,y,x;z)$.

\subsection*{$\mathbf{K_{JJSSJ}}$}
Using eq. (\ref{LLR})  we obtain:
\begin{equation}\begin{split}
&f^{acb}S_A^{dc}(z)S_A^{eb}(z^{\prime})J_{L}^{d}(x)J_{L}^{e}(y)J_{R}^{a}(w)s(u,v)=\frac{i}{4N_{c}}\left[\delta(v-y)-\delta(u-y)\right]\left[\delta(v-w)-\delta(u-w)\right]\\
&\times \left[\left\{ tr\left[S(v)S^{\dagger}(z)S(z^{\prime})S^{\dagger}(u)S(z)S^{\dagger}(z^{\prime})\right]-N_{c}^{3}s(u,z)s(z,z^{\prime})s(z^{\prime},v)\right\} \delta(u-x)\right.\\
&\left.+\left\{ tr\left[S^{\dagger}(u)S(z^{\prime})S^{\dagger}(z)S(v)S^{\dagger}(z^{\prime})S(z)\right]-N_{c}^{3}s(z^{\prime},z)s(z,v)s(u,z^{\prime})\right\} \delta(v-x)\right]\,.
 \end{split}\end{equation}
Similarly 
\begin{equation}\begin{split}
&-f^{acb}S_A^{cd}(z)S_A^{be}(z^{\prime})J_{R}^{d}(x)J_{R}^{e}(y)J_{L}^{a}(w)s(u,v)=\frac{i}{4N_{c}}\left[\delta(v-y)-\delta(u-y)\right]\left[\delta(v-w)-\delta(u-w)\right]\\
&\times \left[\left\{ tr\left[S^{\dagger}(u)S(z)S^{\dagger}(z^{\prime})S(v)S^{\dagger}(z)S(z^{\prime})\right]-N_{c}^{3}s(z,v)s(z^{\prime},z)s(u,z^{\prime})\right\} \delta(v-x)\right.\\
&\left.+\left\{ tr\left[S(v)S^{\dagger}(z^{\prime})S(z)S^{\dagger}(u)S(z^{\prime})S^{\dagger}(z)\right]-N_{c}^{3}s(z,z^{\prime})s(u,z)s(z^{\prime},v)\right\} \delta(u-x)\right]\,. \\
 \end{split}\end{equation}
This results in
 \begin{equation}\begin{split}
&\int_{x,y,w,z,z^{\prime}}K_{JJSSJ}(w,x,y;z,z^{\prime})f^{acb}S_A^{dc}(z)S_A^{eb}(z^{\prime})J_{L}^{d}(x)J_{L}^{e}(y)J_{R}^{a}(w)s(u,v)=\\
&=\int_{z,z^{\prime}}\frac{i}{4N_{c}}\left[tr\left[S(v)S^{\dagger}(z)S(z^{\prime})S^{\dagger}(u)S(z)S^{\dagger}(z^{\prime})\right]-N_{c}^{3}s(u,z)s(z,z^{\prime})s(z^{\prime},v)\right]\\
&\times \left[K_{JJSSJ}(v,u,v;z,z^{\prime})-K_{JJSSJ}(u,u,v;z,z^{\prime})-K_{JJSSJ}(v,u,u;z,z^{\prime})-\right.\\ 
& +K_{JJSSJ}(u,u,u;z,z^{\prime}) -K_{JJSSJ}(v,v,v;z,z^{\prime})+K_{JJSSJ}(u,v,v;z,z^{\prime})\\
&\left.+K_{JJSSJ}(v,u,v;z,z^{\prime})-K_{JJSSJ}(u,u,v;z,z^{\prime})\right]\\
  \end{split}\end{equation}
and:
   \begin{equation}\begin{split}
&-\int_{x,y,w,z,z^{\prime}}K_{JJSSJ}(w,x,y;z,z^{\prime})f^{acb}S_A^{cd}(z)S_A^{be}(z^{\prime})J_{R}^{d}(x)J_{R}^{e}(y)J_{L}^{a}(w)s(u,v)=\\
&\int_{z,z^{\prime}}\frac{i}{4N_{c}}\left[tr\left[S^{\dagger}(u)S(z)S^{\dagger}(z^{\prime})S(v)S^{\dagger}(z)S(z^{\prime})\right]-N_{c}^{3}s(z,v)s(z^{\prime},z)s(u,z^{\prime})\right]\times\\
&\left[K_{JJSSJ}(u,v,u;z,z^{\prime})-K_{JJSSJ}(v,v,u;z,z^{\prime})-K_{JJSSJ}(u,v,v;z,z^{\prime})\right.\\
&+K_{JJSSJ}(v,v,v;z,z^{\prime})-K_{JJSSJ}(u,u,u;z,z^{\prime})+K_{JJSSJ}(v,u,u;z,z^{\prime})\\
&\left. +K_{JJSSJ}(u,v,u;z,z^{\prime})-K_{JJSSJ}(v,v,u;z,z^{\prime})\right]\,.\\
     \end{split}\end{equation}
Combining the two terms and using the antisymmetry of the kernel under 
simultaneous interchange of $x\leftrightarrow y$ and $z\leftrightarrow  z^\prime$:  $K_{JJSSJ}(w;x,y;z^{\prime},z)=-K_{JJSSJ}(w;y,x,z,z^{\prime})$
we obtain the final expression:
   \begin{eqnarray}
&&\int_{w,x,y,z,z^{\prime}}K_{JJSSJ}(w;x,y;z,z^{\prime})f^{acb}\nonumber \\
&&\ \ \ \times \left[J_{L}^{d}(x)J_{L}^{e}(y)S_A^{dc}(z)S_A^{eb}(z^{\prime})J_{R}^{a}(w)-J_{L}^{a}(w)S_A^{cd}(z)S_A^{be}(z^{\prime})J_{R}^{d}(x)J_{R}^{e}(y)\right]s(u,v)\nonumber 
\\
&&=-\frac{i}{4N_{c}}\int_{z,z^{\prime}}\left[4K_{JJSSJ}(u;v,u;z,z^{\prime})-4K_{JJSSJ}(v;v,u;z,z^{\prime})-2K_{JJSSJ}(u;v,v;z,z^{\prime})\right. \nonumber \\
&&\left. \ \ \ + 2K_{JJSSJ}(v;v,v;z,z^{\prime})-2K_{JJSSJ}(u;u,u;z,z^{\prime})+2K_{JJSSJ}(v;u,u;z,z^{\prime})\right]\nonumber \\
&&\ \ \ \times N_{c}^{3}s(z,v)s(z^{\prime},z)s(u,z^{\prime})\nonumber \\
&&\ \ \ +\frac{i}{4N_{c}}\int_{z,z^{\prime}}\left[2K_{JJSSJ}(v;u,v;z,z^{\prime})-2K_{JJSSJ}(u;u,v;z,z^{\prime})+2K_{JJSSJ}(u;v,u;z,z^{\prime})\right.\nonumber \\
&&\ \ \ \left. -2K_{JJSSJ}(v;v,u;z,z^{\prime})\right]
tr\left[S(v)S^{\dagger}(z)S(z^{\prime})S^{\dagger}(u)S(z)S^{\dagger}(z^{\prime})\right]\nonumber \\
&&=\frac{i}{2N_{c}}\int_{z,z^{\prime}}\left[K_{JJSSJ}(u;u,u;z,z^{\prime})-K_{JJSSJ}(u;v,u;z,z^{\prime})+K_{JJSSJ}(u;v,v;z,z^{\prime})\right. \nonumber \\
&& \ \ \ -K_{JJSSJ}(u;u,v;z,z^{\prime})+K_{JJSSJ}(v;u,v;z,z^{\prime})-K_{JJSSJ}(v;u,u;z,z^{\prime})\nonumber \\
&&\left. \ \ \ +K_{JJSSJ}(v;v,u,z,z^{\prime})-K_{JJSSJ}(v;v,v;z,z^{\prime})\right]
N_{c}^{3}s(z,v)s(z^{\prime},z)s(u,z^{\prime})\nonumber \\
&&\ \ \ +\frac{1}{N_{c}}\int_{z,z^{\prime}}\widetilde{K}(u,v,z,z^{\prime})\times\left[N_{c}^{3}s(z,v)s(z^{\prime},z)s(u,z^{\prime})-tr\left[S(v)S^{\dagger}(z)S(z^{\prime})S^{\dagger}(u)S(z)S^{\dagger}(z^{\prime})\right]\right]\, , \nonumber \\
\end{eqnarray}
where $\tilde K$ is defined in (\ref{tildeK}).
For the virtual terms, using (\ref{LLR})  we find:
\begin{eqnarray}
&&\frac{1}{3}\int_{w,x,y,z,z^{\prime}}K_{JJSSJ}(w;x,y;z,z^{\prime})f^{acb}\left[J_{L}^{c}(x)J_{L}^{b}(y)J_{L}^{a}(w)-J_{R}^{c}(x)J_{R}^{b}(y)J_{R}^{a}(w)\right]s(u,v)\nonumber \\
&&=-\int_{z,z^{\prime}}\left[-K_{JJSSJ}(u;v,v;z,z^{\prime})+K_{JJSSJ}(u;u,v;z,z^{\prime})+K_{JJSSJ}(v;v,v;z,z^{\prime})\right.\nonumber \\
&&-K_{JJSSJ}(v;u,v;z,z^{\prime})+K_{JJSSJ}(v;v,u;z,z^{\prime})-K_{JJSSJ}(v;u,u;z,z^{\prime})\nonumber \\
&&\left.+K_{JJSSJ}(u;u,u;z,z^{\prime})-K_{JJSSJ}(u;v,u;z,z^{\prime})\right]\frac{N_{c}^{2}-1}{6}is(u,v)\nonumber\\
&&=-\frac{(N_{c}^{2}-1)}{3}\,\int_{z,z^{\prime}}\widetilde{K}(u,v,z,z^{\prime})s(u,v)\,.
  \end{eqnarray}

\subsection*{Putting all the pieces together}
Combining all the terms together, we can finally write down an evolution equation of the dipole $s$ at NLO, expressed in terms of the kernels parametrizing
the Hamiltonian. 
\begin{eqnarray}\label{ourdipole}
&&-\frac{d}{dY}s(u,v)  =\nonumber \\
&&=\int_{z}\left[2N_{c}K_{JSJ}(u,v;z)-iN_{c}^{2}\left(K_{JJSJ}(v,u,v,z)+K_{JJSJ}(u,v,u,z)\right)+N_{c}^{2}\int_{z^{\prime}}\widetilde{K}(u,v,z,z^{\prime})\right]\nonumber\\
&&\times \left[s(u,z)s(z,v)-s(u,v)\right]\nonumber\\
&&-\frac{4}{N_{c}}\int_{z\, z^{\prime}}\, K_{q\bar{q}}(u,v;z,z^{\prime})\left[tr\left[S^{\dagger}(u)t^{a}S(v)t^{b}\right]tr\left[S^{\dagger}(z)t^{a}S(z^{\prime})t^{b}\right]-(z\rightarrow z^{\prime})\right]\nonumber \\
&&-\frac{1}{N_{c}}\int_{z,z^{\prime}}\,\left[K_{JSSJ}(u,v;z,z^{\prime})-\widetilde{K}(u,v,z,z^{\prime})\right]
\left[N_{c}^{3}s(u,z^{\prime})s(z^{\prime},z)s(z,v)\right.\nonumber \\
&&\left. -tr\left[S(v)S^{\dagger}(z)S(z^{\prime})S^{\dagger}(u)S(z)S^{\dagger}(z^{\prime})\right]-N_{c}^{3}s(u,z)s(z,v)+N_{c}s(u,v)\right]\nonumber \\
&&+\frac{i}{2N_{c}}\int_{z,z^{\prime}}\left[K_{JJSSJ}(u,u,u,z,z^{\prime})-K_{JJSSJ}(u,v,u,z,z^{\prime})+K_{JJSSJ}(u,v,v,z,z^{\prime})\right. \nonumber \\
&&-K_{JJSSJ}(u,u,v,z,z^{\prime})+K_{JJSSJ}(v,u,v,z,z^{\prime})-K_{JJSSJ}(v,u,u,z,z^{\prime})\nonumber \\
&&\left.+K_{JJSSJ}(v,v,u,z,z^{\prime})-K_{JJSSJ}(v,v,v,z,z^{\prime})\right] \,N_{c}^{3}s(z,v)s(z^{\prime},z)s(u,z^{\prime})\,.
\end{eqnarray}
We used the relation eq.(\ref{Krelate}) to simplify the virtual term in the first line.


 \subsection*{Determination of the kernels}

To determine the kernels we compare the dipole evolution (\ref{dipole}) of ref. \cite{BC} with the one
derived in (\ref{ourdipole}). Recall that the kernels $K_{JJSSJ}$ and $K_{JJSJ}$ are known from comparison with the results of \cite{Grab} and are given in
eq.(\ref{KJJSJ}). Both kernels $K_{JJSSJ}$ and $K_{JJSJ}$ contribute to the evolution of the dipole $s$ and their contributions constitute a non-trivial consistency check. 

The strategy of comparison between (\ref{dipole})  in (\ref{ourdipole}) is straightforward: the right hand sides of the evolution equations
contain various independent operators and we have to match their respective coefficients. 

The simplest example of the procedure is the "quark term"
identified by the operator 
$tr\left[S^{\dagger}(u)t^{a}S(v)t^{b}\right]tr\left[S^{\dagger}(z)t^{a}S(z^{\prime})t^{b}\right]$ and its subtraction.
Comparing the coefficients we immediately read off the kernel $K_{qq}$ as quoted in (\ref{Kqq}).

Next consider the coefficient of the operator $s(u,z)s(z,v)-s(u,v)$. We first notice that,  given the expression  (\ref{KJJSJ}) for the kernel $K_{JJSJ}$, the combination 
\begin{equation}\begin{split}\label{loglog}
&iN_{c}^{2}\left[K_{JJSJ}(v,u,v,z)+K_{JJSJ}(u,v,u,z)\right]\\
&\ \ \ \ \ =\frac{\alpha_{s}^{2}N_{c}^{2}}{4\pi^{3}}\left[\frac{2U\cdot V}{U^{2}V^{2}}-\frac{V^{2}}{V^{4}}-\frac{U^{2}}{U^{4}}\right]\log\frac{V^{2}}{(u-v)^{2}}\log\frac{U^{2}}{(u-v)^{2}}\\
&\ \ \ \ \  =-\frac{\alpha_{s}^{2}N_{c}^{2}}{4\pi^{3}}\frac{(u-v)^{2}}{U^{2}V^{2}}\log\frac{V^{2}}{(u-v)^{2}}\log\frac{U^{2}}{(u-v)^{2}}\\
\end{split}\end{equation}
reproduces exactly the corresponding $\ln\ln$ term in (\ref{dipole}). The rest of the coefficient determines the kernel $K_{JSJ}$ as quoted in (\ref{KJSJ}).
Similarly comparison of the coefficient of\\  $N_{c}^{3}s(u,z)s(z,z^{\prime})s(z^{\prime},v)-tr\left[S(v)S^{\dagger}(z)S(z^{\prime})S^{\dagger}(u)S(z)S^{\dagger}(z^{\prime})\right]-(z^{\prime}\rightarrow z)$ determines $K_{JSSJ}$ (\ref{KJSSJ}).

 
 Lastly consider the coefficient of the operator $s(z,v)s(z^{\prime},z)s(u,z^{\prime})$. In (\ref{ourdipole}) this coefficient is given by in terms of a sum of several $K_{JJSSJ}$ kernels.
 Thus we have to demonstrate that our expression for $K_{JJSSJ}(w;x,y,z,z^{\prime})$ as given by (\ref{KJJSJ}) indeed reproduces the corresponding
coefficient in  (\ref{dipole}).
 To this end consider the combination:
\begin{eqnarray}
&&K_{JJSSJ}(v,v,u,z,z^{\prime})-K_{JJSSJ}(v,u,u,z,z^{\prime})-K_{JJSSJ}(v,v,v,z,z^{\prime})+K_{JJSSJ}(v,u,v,z,z^{\prime})\nonumber \\
&&=-\frac{i\alpha_{s}^{2}}{2\pi^{4}}\left[\frac{\delta_{ij}}{2(z-z^{\prime})^{2}}-\frac{(z-z^{\prime})_{i}V_{j}^{\prime}}{(z-z^{\prime})^{2}(V^{\prime})^{2}}+\frac{(z-z^{\prime})_{j}V_{i}}{(z-z^{\prime})^{2}V^{2}}-\frac{V_{i}V_{j}^{\prime}}{V^{2}(V^{\prime})^{2}}\right]\nonumber \\
&&\ \ \ \ \ \times \left[\frac{U_{i}V_{j}^{\prime}}{U^{2}(V^{\prime})^{2}}+\frac{V_{i}U_{j}^{\prime}}{V^{2}(U^{\prime})^{2}}-\frac{V_{i}V_{j}^{\prime}}{V^{2}(V^{\prime})^{2}}-\frac{U_{i}U_{j}^{\prime}}{U^{2}(U^{\prime})^{2}}\right]ln\frac{V^{2}}{(V^{\prime})^{2}}\,.
 \end{eqnarray}
Some algebraic manipulations lead to
\begin{equation}\begin{split}
&\frac{\delta_{ij}}{2(z-z^{\prime})^{2}}\left[\frac{U_{i}V_{j}^{\prime}}{U^{2}(V^{\prime})^{2}}+\frac{V_{i}U_{j}^{\prime}}{V^{2}(U^{\prime})^{2}}-\frac{V_{i}V_{j}^{\prime}}{V^{2}(V^{\prime})^{2}}-\frac{U_{i}U_{j}^{\prime}}{U^{2}(U^{\prime})^{2}}\right]=\\
&\frac{1}{(z-z^{\prime})^{2}}\left[\frac{UV^{\prime}}{2U^{2}(V^{\prime})^{2}}+\frac{VU^{\prime}}{2V^{2}(U^{\prime})^{2}}-\frac{VV^{\prime}}{2V^{2}(V^{\prime})^{2}}-\frac{UU^{\prime}}{2U^{2}(U^{\prime})^{2}}\right]\\\\
&-\frac{(z-z^{\prime})_{i}V_{j}^{\prime}}{(z-z^{\prime})^{2}(V^{\prime})^{2}}\left[\frac{U_{i}V_{j}^{\prime}}{U^{2}(V^{\prime})^{2}}+\frac{V_{i}U_{j}^{\prime}}{V^{2}(U^{\prime})^{2}}-\frac{V_{i}V_{j}^{\prime}}{V^{2}(V^{\prime})^{2}}-\frac{U_{i}U_{j}^{\prime}}{U^{2}(U^{\prime})^{2}}\right]=\\
&-\frac{1}{(z-z^{\prime})^{2}}\left[\frac{(z-z^{\prime})U}{(V^{\prime})^{2}U^{2}}+\frac{\left[(z-z^{\prime})V\right]V^{\prime}U^{\prime}}{(V^{\prime})^{2}V^{2}(U^{\prime})^{2}}-\frac{\left[(z-z^{\prime})V\right]}{V^{2}(V^{\prime})^{2}}-\frac{\left[(z-z^{\prime})U\right]V^{\prime}U^{\prime}}{(V^{\prime})^{2}U^{2}(U^{\prime})^{2}}\right]\\\\
&\frac{(z-z^{\prime})_{j}V_{i}}{(z-z^{\prime})^{2}V^{2}}\left[\frac{U_{i}V_{j}^{\prime}}{U^{2}(V^{\prime})^{2}}+\frac{V_{i}U_{j}^{\prime}}{V^{2}(U^{\prime})^{2}}-\frac{V_{i}V_{j}^{\prime}}{V^{2}(V^{\prime})^{2}}-\frac{U_{i}U_{j}^{\prime}}{U^{2}(U^{\prime})^{2}}\right]=\\
&\frac{1}{(z-z^{\prime})^{2}}\left[\frac{\left[(z-z^{\prime})V^{\prime}\right]VU}{V^{2}U^{2}(V^{\prime})^{2}}+\frac{(z-z^{\prime})U^{\prime}}{V^{2}(U^{\prime})^{2}}-\frac{(z-z^{\prime})V^{\prime}}{V^{2}(V^{\prime})^{2}}-\frac{\left[(z-z^{\prime})U^{\prime}\right]VU}{V^{2}U^{2}(U^{\prime})^{2}}\right]\\\\
&-\frac{V_{i}V_{j}^{\prime}}{V^{2}(V^{\prime})^{2}}\left[\frac{U_{i}V_{j}^{\prime}}{U^{2}(V^{\prime})^{2}}+\frac{V_{i}U_{j}^{\prime}}{V^{2}(U^{\prime})^{2}}-\frac{V_{i}V_{j}^{\prime}}{V^{2}(V^{\prime})^{2}}-\frac{U_{i}U_{j}^{\prime}}{U^{2}(U^{\prime})^{2}}\right]\\
&=-\frac{1}{V^{2}(V^{\prime})^{2}}\left[\frac{VU}{U^{2}}+\frac{V^{\prime}U^{\prime}}{(U^{\prime})^{2}}-1-\frac{\left[VU\right]V^{\prime}U^{\prime}}{U^{2}(U^{\prime})^{2}}\right]\,. \\
 \end{split}\end{equation}
Finally ( defining $Z=z-z^{\prime}$) we obtain:
\begin{eqnarray}
K_{JJSSJ}(v,u,u,z,z^{\prime})-K_{JJSSJ}(v,v,u,z,z^{\prime})+K_{JJSSJ}(v,v,v,z,z^{\prime})-K_{JJSSJ}(v,u,v,z,z^{\prime})\nonumber \\
=\frac{i\alpha_{s}^{2}}{8\pi^{4}}\frac{(u-v)^{2}}{U^{2}V^{2}(U^{\prime})^{2}(V^{\prime})^{2}}\Big[\frac{1}{Z^{2}}(2U^{2}V^{2}+U^{2}Z^{2}+V^{2}Z^{2}+2V^{2}(U\cdot Z)
\nonumber \\
 +2U^{2}(V\cdot Z))-(u-v)^{2}\Big] \ln\left(\frac{V^{2}}{(V^{\prime})^{2}}\right)\,. \nonumber \\
 \end{eqnarray}
Similarly:
\begin{eqnarray}
K_{JJSSJ}(u,u,u,z,z^{\prime})-K_{JJSSJ}(u,v,u,z,z^{\prime})+K_{JJSSJ}(u,v,v,z,z^{\prime})-K_{JJSSJ}(u,u,v,z,z^{\prime})\nonumber \\
=\frac{i\alpha_{s}^{2}}{8\pi^{4}}\frac{(u-v)^{2}}{U^{2}V^{2}(U^{\prime})^{2}(V^{\prime})^{2}}\Big[\frac{1}{Z^{2}}(2U^{2}V^{2}+U^{2}Z^{2}+V^{2}Z^{2}+2V^{2}(U\cdot Z)
\nonumber \\ +2U^{2}(V\cdot Z))-(u-v)^{2}\Big]\ln\left(\frac{U^{2}}{(U^{\prime})^{2}}\right)\,. \nonumber \\
\end{eqnarray}
Combining the last two expressions yields the coefficient of the $s(z,v)s(z^{\prime},z)s(u,z^{\prime})$ term:
\begin{eqnarray}
&&-\frac{\alpha_{s}^{2}}{16N_{c}\pi^{4}}
\frac{(u-v)^{2}}{U^{2}V^{2}(U^{\prime})^{2}(V^{\prime})^{2}}\Big[\frac{1}{Z^{2}}(2U^{2}V^{2}+U^{2}Z^{2}+V^{2}Z^{2}+2V^{2}(U\cdot Z)\nonumber \\
&&\ \ \ \ \ \ \ \ \ \ +2U^{2}(V\cdot Z))-(u-v)^{2}\Big] \ln\left(\frac{U^{2}(V^{\prime})^{2}}{(U^{\prime})^{2}V^{2}}\right)N_{c}^{3}\nonumber \\
&&=-\frac{\alpha_{s}^{2}}{16N_{c}\pi^{4}}
\left\{ \frac{(u-v)^{2}}{(z-z^{\prime})^{2}}\left[\frac{1}{U^{2}(V^{\prime})^{2}}+\frac{1}{V^{2}(U^{\prime})^{2}}\right]-\frac{(u-v)^{4}}{U^{2}V^{2}(U^{\prime})^{2}(V^{\prime})^{2}}\right\} \ln\left(\frac{U^{2}(V^{\prime})^{2}}{(U^{\prime})^{2}V^{2}}\right)N_{c}^{3}\nonumber \\
\end{eqnarray}
which reproduces the appropriate coefficient in eq.(\ref{dipole}).


 \section{Useful identities}
 In this appendix we list some algebraic identities that we found useful in this paper.
 
\begin{equation}
S_{A}^{ab}(x)(t^{a}S(x)t^{b})_{ij}
=(t^{a}t^{a}S(x))_{ij}\,.
\end{equation}
Also
\begin{equation}
S_{A}^{ab}(x)(t^{a}S(x))_{ij}(S(x)t^{b})_{kl}
=(t^{a}S(x))_{ij}(t^{a}S(x))_{sl}\,.
\end{equation}
Therefore we can replace:
\begin{equation}\begin{split}
&J_{L}^{a}(x)\, J_{L}^{a}(y)+J_{R}^{a}(x)\, J_{R}^{a}(y)-2\, J_{L}^{a}(x)\, S_{A}^{ab}(z)\, J_{R}^{b}(y)\\
&=\left[S_{A}^{ab}(x)+S_{A}^{ab}(y)-2S_{A}^{ab}(z)\right]\, J_{L}^{a}(x)\, J_{R}^{b}(y)\,.
\end{split}\end{equation}
Another identity 
\begin{equation}
f^{ade}f^{bd^{\prime}e^{\prime}}S_{A}^{dd^{\prime}}(w)S_{A}^{ee^{\prime}}(w)=
N_{c}S_{A}^{ab}(w)\,.
  \end{equation}
This leads to
\begin{equation}\begin{split}
&f^{abc}f^{def}\, J_{L}^{a}(x)\, S_{A}^{be}(z)\, S_{A}^{cf}(v)\, J_{R}^{d}(y)-N_{c}\, J_{L}^{a}(x)\, S_{A}^{ab}(z)\, J_{R}^{b}(y)\\
&= f^{adc}f^{bef}\left[S_{A}^{de}(z)\, S_{A}^{cf}(v)-S_{A}^{de}(z)\, S_{A}^{cf}(z)\right]\, J_{L}^{a}(x)\, J_{R}^{b}(y)\,. \\
   \end{split}\end{equation}
\begin{equation}\begin{split}\label{id1}
&\int_{z,z^{\prime}}S_{A}^{dd^{\prime}}(z)\left[S_{A}^{ee^{\prime}}(z^{\prime})-S_{A}^{ee^{\prime}}(z)\right]\frac{X\cdot X^{\prime}}{(z-z^{\prime})^{2}X^{2}(X^{\prime})^{2}}\ln\frac{X^{2}}{(X^{\prime})^{2}}\\
&\times \left[f^{ad^{\prime}e^{\prime}}(\{t^{d},t^{e}\}S(x)t^{a})_{ij}-f^{ade}(t^{a}S(x)\{t^{d^{\prime}},t^{e^{\prime}}\})_{ij}\right]\\
&=\int_{z,z^{\prime}}S_{A}^{dd^{\prime}}(z)\left[S_{A}^{ee^{\prime}}(z^{\prime})-S_{A}^{ee^{\prime}}(z)\right]\frac{X\cdot X^{\prime}}{(z-z^{\prime})^{2}X^{2}(X^{\prime})^{2}}\ln\frac{X^{2}}{(X^{\prime})^{2}}\\ 
&\times \left[f^{ad^{\prime}e^{\prime}}(t^{d}t^{e}S(x)t^{a})_{ij}-f^{ade}(t^{a}S(x)t^{d^{\prime}}t^{e^{\prime}})_{ij}\right]\\
&+\int_{z,z^{\prime}}S_{A}^{dd^{\prime}}(z)\left[S_{A}^{ee^{\prime}}(z^{\prime})+S_{A}^{ee^{\prime}}(z)\right]\frac{X\cdot X^{\prime}}{(z-z^{\prime})^{2}X^{2}(X^{\prime})^{2}}\ln\frac{X^{2}}{(X^{\prime})^{2}}\\ 
&\times \left[f^{ad^{\prime}e^{\prime}}(t^{d}t^{e}S(x)t^{a})_{ij}-f^{ade}(t^{a}S(x)t^{d^{\prime}}t^{e^{\prime}})_{ij}\right]\\
&=2\int_{z,z^{\prime}}S_{A}^{dd^{\prime}}(z)S_{A}^{ee^{\prime}}(z^{\prime})\frac{X\cdot X^{\prime}}{(z-z^{\prime})^{2}X^{2}(X^{\prime})^{2}}\ln\frac{X^{2}}{(X^{\prime})^{2}}\left[f^{ad^{\prime}e^{\prime}}(t^{d}t^{e}S(x)t^{a})_{ij}-f^{ade}(t^{a}S(x)t^{d^{\prime}}t^{e^{\prime}})_{ij}\right]\\
&=2\int_{z,z^{\prime}}S_{A}^{dc}(z)S_{A}^{eb}(z^{\prime})\frac{X\cdot X^{\prime}}{(z-z^{\prime})^{2}X^{2}(X^{\prime})^{2}}\ln\frac{X^{2}}{(X^{\prime})^{2}}\left[f^{acb}(t^{e}t^{d}S(x)t^{a})_{ij}-f^{ade}(t^{a}S(x)t^{c}t^{b})_{ij}\right]
\\
&+2\int_{z,z^{\prime}}S_{A}^{dc}(z)S_{A}^{eb}(z^{\prime})\frac{X\cdot X^{\prime}}{(z-z^{\prime})^{2}X^{2}(X^{\prime})^{2}}\ln\frac{X^{2}}{(X^{\prime})^{2}}f^{acb}f^{deh}(t^{h}S(x)t^{a})_{ij}\,. 
 \end{split}\end{equation}
The last term vanishes due to the anti-symmetry under exchanging $z$ and $z^{\prime}$.


\section*{Acknowledgments}
We are most grateful to Ian Balitsky who inspired us for this project.  M.L and Y.M. thank the Physics Department of the University of Connecticut for hospitality
when this project was initiated.
The research was supported by the DOE grant DE-FG02-13ER41989; the EU FP7 grant PIRG-GA-2009-256313; the 
 ISRAELI SCIENCE FOUNDATION grant \#87277111;   the People Program (Marie Curie Actions) of the European Union's Seventh Framework   under REA
grant agreement \#318921;  and the BSF grant \#2012124.

\end{document}